\documentclass[onecolumn]{aa} 
\pdfoutput=1
\usepackage{latexsym}
\usepackage{amssymb}
\usepackage{graphicx}
\usepackage{lscape}
\usepackage{color}
\usepackage[authoryear]{natbib}
\bibpunct{(}{)}{;}{a}{}{,} 
\usepackage{txfonts}
\def\av{A$_{\rm v}$}

\def\micron{$\mu$m}
\def\msun{\ifmmode M_{\odot} \else M$_{\odot}$\fi}
\def\msunyr{\ifmmode M_{\odot} {\rm yr}^{-1} \else M$_{\odot}$ yr$^{-1}$\fi}
\def\hyperz{{\em Hyperz}}
\def\flyf{\ifmmode f_{\rm Lyf} \else $f_{\rm Lyf}$\fi}
\def\pz{\ifmmode P(z) \else $P(z)$\fi}
\def\ki2{\ifmmode \chi^2 \else $\chi^2$\fi}
\def\zphot{\ifmmode z_{\rm phot} \else $z_{\rm phot}$\fi}
\def\zfit{\ifmmode z_{\rm fit} \else $z_{\rm fit}$\fi}
\def\lsun{\ifmmode L_{\odot} \else L$_{\odot}$\fi}
\def\lbol{\ifmmode L_{\rm bol} \else L$_{\rm bol}$\fi}

\begin{document}
   \title{EROs found behind lensing clusters}
   \subtitle{II. Empirical properties, classification, and SED modelling
based on multi-wavelength observations 
\thanks{Based on observations collected at the Very Large Telescope (Antu/UT1), 
European Southern Observatory, Paranal, Chile 
(ESO Programs 69.A-0508, 70.A-0355, 73.A-0471), 
the NASA/ESA \textit{Hubble Space Telescope} obtained at the Space Telescope 
Science Institute which is operated by AURA under NASA contract NAS5-26555,
the Spitzer Space Telescope, which is operated by the Jet Propulsion Laboratory, 
California Institute of Technology under NASA contract 1407,
and the Chandra satellite.}}

\author{A. Hempel \inst{1} \and D. Schaerer\inst{1,3} \and
  E. Egami\inst{2}\and R. Pell\'{o}\inst{3} \and  M. Wise\inst{4} \and
  J. Richard\inst{5,3}  \and  J.-F. Le Borgne\inst{3} \and J.-P. Kneib\inst{6} }

   \institute{Observatoire de Gen\`{e}ve, 51, chemin des Maillettes, CH-1290 Sauverny, Switzerland
              \email{angela.hempel@obs.unige.ch}
              \email{daniel.schaerer@obs.unige.ch}
         \and
          Steward Observatory, University of Arizona, 933 North Cherry Street, Tucson, AZ 85721,USA
         \and
             Observatoire Midi-Pyr\'{e}n\'{e}es, Laboratoire
             d`Astrophysique, UMR 5572, 14 Avenue E.Belin, F-31400
             Toulouse, France
         \and
          Astronomical Institute Anton Pannekoek, Kruislaan 403,
          NL-1098 SJ Amsterdam, The Netherlands
         \and
          Caltech Astronomy, MC105-24, Pasadena, CA 91125, USA
         \and
          OAMP, Laboratoire d'Astrophysique de Marseille, UMR 6110 traverse du Siphon, 
		F-13012 Marseille, France
             }

\date{Received date / Accepted date}

  \abstract
   {}
  {We study the properties and nature of extremely red galaxies (ERO, $R-K\ge$5.6) found behind
two lensing clusters and compare them with other known galaxy populations.}
   {New HST/ACS observations, Spitzer IRAC and MIPS, and Chandra/ACIS observations
of the two lensing clusters Abell 1835 and AC114 contemplate 
our earlier optical and near-IR observations 
 (Richard et al. 2006) and have been used to study extremely red objects (EROs) in these 
deep fields.}
   {We have found 6 and 9 EROs in Abell 1835 and AC114.
Several (7) of these objects are undetected up to the $I$ and/or $z_{850}$
band, and are hence ``optical'' drop-out sources. The photometric redshifts 
of most of our sources (80\%) are $z\sim$ 0.7--1.5. 
According to simple colour-colour diagrams the majority of
our objects would be classified as hosting old stellar populations (``ellipticals'').
However, there are clear signs of dusty starbursts for several among them.
These objects correspond to the most extreme ones in $R-K$ colour. 
We estimate a surface density
of (0.97$\pm$0.31) arcmin$^{-2}$ for EROs with ($R-K \ge $5.6) at
K$<20.5$. 
Among our 15 EROs  6 (40 \%) also classify as distant red galaxies (DRGs).
11 of 13 EROs (85 \%) with available IRAC photometry also fulfil
the selection criteria for IRAC selected EROs (IEROs) of Yan et
al.\ (2004). SED modelling shows that $\sim$ 36 \% of the IEROs in our
sample are luminous or ultra-luminous infrared galaxies ((U)LIRG).
Some very red DRGs are found to be very dusty starbursts, even (U)LIRGs, 
as also supported by their mid-IR photometry. 
No indication for AGNs is found, although faint activity cannot be excluded
for all objects. From mid-IR and X-ray data 5 objects are clearly classified 
as starbursts. 
The derived properties are quite similar 
to those of DRGs and IEROs, except for 5 extreme objects in terms of colours,
for which a very high extinction ($A_V \ga 3$) is found.
}
   {}


   \maketitle
%

\section{Introduction}

Since their discovery in the late 1980ies \citep{elston88,elston89}, extremely red objects (EROs) 
have attracted serious attention. 
These first detections were initially presumed to be high-redshift ($z>6$) 
galaxies in a star-forming phase 
\citep{elston88}. Multi-colour follow-up observations later identified these 
objects as luminous galaxies at $z=0.8$, dominated by an old stellar population 
\citep{elston89}. The detection of two bright (K$\ga$18.4) extended objects 
(HR10 \& HR14) with ($I-K$) colours near 6.5 by Hu and Ridgway (\citeyear{hu94}) 
highlighted the difficulty in classifying these galaxies. When first discovered, 
HR10 and HR14 were interpreted as being ellipticals at $z \sim 2.4$. Subsequent 
spectroscopic and morphological observations indicated that HR10 is not a quiescent 
elliptical galaxy, but rather a bright interacting galaxy at $z=1.44$
\citep{graham96, stern2006}.\\
In general there are two main scenarios which would produce a red
enough spectral energy distribution to satisfy the established colour
criteria for EROs (e.g. $R-K>$5-7, $I-K>$4-6), either due to an old
passively evolved population or by extreme dust reddening as found in
star bursts \citep{cowie94,cimatti99,daddi02,georga06} in a redshift
range of $1 \la z \la2$. A number of review articles discuss
various aspects related to this topic \citep[e.g.][]{mccarthy2004,ferguson2000}.  \\
Especially the abundance of massive old ellipticals poses a strong test for 
the two competing scenarios of elliptical galaxy formation: early
assembly ($z_{f}>$2-3), e.g. by monolithic collapse, and passive luminosity 
evolution thereafter (PLE models) \citep{tinsley76,pozzetti96}, or 
hierarchical merging of smaller sized objects \citep{white78,kauffmann93,somerville01}.
Observational evidence has been found for both scenarios: several
surveys detected a deficit of ellipticals at $z>$1, supporting the
hierarchical merging models \citep{roche03,kitzbichler06}, while
others are consistent with PLE
\citep{im02,cimatti02b,somerville04}. \\
However, in recent years the
hierarchical merging scenario in a $\Lambda$CDM universe has been
established as the favoured model. Nevertheless, the vast number of
different renditions leaves room for dramatically different
predictions regarding critical parameters like the number density of
massive galaxies at specific times \citep[and references
therein]{fontana04,treu05}.\\
The picture is complicated by the results of numerous morphological
studies on EROs, which assigns a large fraction of EROs to disk
galaxies  at somewhat lower redshifts \citep{yan03, gilbank03, moustakas04}. 
In addition, a small fraction of EROs could also be active galactic nuclei (AGNs), 
as shown by deep XMM and Chandra data
\citep{alexander02,roche03,brusa2005}. \\
With the increasing number of large scale surveys like UKIDSS
\citep{simpson2006} and others it became clear that other means than
spectroscopy is needed in order to classify to which of the two major
galaxy populations the large number of EROs belong. These could either 
be combinations of
$RIJHK$ \citep{pozzetti00,wiklind04} colours, or the use of near and
mid-infrared bands \citep{wilson04}. \\
Independently of which exact colour criteria has been used, all EROs
have at least one mutual property, their faintness at optical
wavelengths causing limitations to the accuracy of photometric redshift 
estimates and other parameters derived from SED features. 
In order to increase the apparent brightness of EROs we use the natural magnification effect
provided by massive galaxy clusters. This method has been applied
successfully for the investigation of other faint sources, like Lyman
break galaxies \citep{pettini2000,swinbank2007}, faint SCUBA sources
\citep{smail1998,ivison2001} and EROs \citep{smith2002,takata03}.

In the present study we proceed to a systematic selection of EROs
in the fields of the two lensing clusters Abell 1835 and AC114,
based on observations obtained by \citet{r06} and
new ACS/HST, Spitzer, and  Chandra observations. 
These are used to discuss their empirical properties, their nature and
to derive physical parameters like photometric redshift, extinction,
star formation rates, and stellar population properties.
Several of these objects were found earlier in our $H$-band selected sample 
of optical drop-out objects
constructed for the search of high redshift galaxies \citep[see][]{r06}.
The ERO subsample from that paper is analysed in detail 
in \citet{s07}.

Throughout this paper we adopted the following cosmology:
$\Omega_{m}=0.3$, H$_{o}$=70 km s$^{-1}$Mpc$^{-1}$ in a flat
universe. All magnitudes are given in the Vega system if not stated otherwise.


\section{Observations and data reduction}
\label{data}
The observations described here are part of multi-colour observations
on two galaxy clusters, AC114 and Abell 1835, which have well known
lensing properties. An
extensive description of the initial observations (optical and
near-infrared data) and the available
data can be found in Richard et al. \citeyearpar{r06}. Exposure time, limiting
magnitudes and more characteristics can be found in
Tab. \ref{tab_data}.\\
The near-infrared data ($SZ,J,H$ and $K_{s}$) were obtained with the Infrared Spectrometer
and Array Camera (ISAAC) located on the VLT-UT1 (FOV 2.5 acrmin x 2.5
arcmin with a pixel size of 0.148 arcsec). The optical data for
Abell 1835 ($VRI$) are archive data from the CFHT12k camera at CFHT \citep{czoske2003}, those for AC114 ($UBVRI$) were taken from \citet{campusano2001}.

\begin{table*}[htb]
\caption{Characteristics of the observations. Filters, effective wavelength, exposure time, 
depth (1$\sigma$), 
and AB corrections ($C_{AB}$) correspond to $m_{AB}=m_{\rm vega} +C_{AB}$
are listed.
$^{a}$ If only one filter name is given, than the same setup was used 
for both fields. $^{b}$ For small fractions of the final $R_{702}$ image, the total
exposure time is only 400 seconds. For objects without R-band
  detection we adjusted the detection limits accordingly (see Tab.\ref{taba1835}).
$^\star$ The 24 $\mu$m integration time is as much as 3600 s within a 
$\sim$30\arcsec\-wide strip crossing the cluster center because of the way multiple data sets were taken.}
\begin{tabular}{ccrrrrrrcc}
\hline
\multicolumn{2}{ c }{~~~~Filter$^{a}$~~~~} & \multicolumn{2}{c }{~~~~$\lambda_{eff}$[nm]}~~~~ & \multicolumn{2}{c }{~~~~$t_{exp}$[sec]~~~~} & \multicolumn{2}{c }{~~~~depth  [mag]~~~~}&\multicolumn{2}{c }{~~~~$C_{AB}$[mag]~~~~}\\
Abell 1835 & AC114 & Abell 1835 & AC114 & Abell 1835 & AC114 & Abell 1835 & AC114 & Abell 1835 & AC114\\
\hline 
     &   &    &    &    &   &    &    &    &   \\          
        &    $U$                  &         & 365               &                & 20000~~~          &       & 29.1     &           &  0.693   \\
        &    $B$                  &         & 443               &                &  9000~~~          &       & 29.0     &           & -0.064   \\
\multicolumn{2}{ c }{$V$}          &  543    & 547               & 3750~~~        & 20000~~~          & 28.1  & 28.5     & 0.018     &  0.022   \\ 
 $R$      & $R_{702}$              &  664    & 700               & 5400~~~        &  8300 (400)$^{b}$ & 27.8  & 27.7 (26.1)$^{b}$ & 0.246     & 0.299   \\ 
 $I$      & $I_{814}$              &  817    & 801               & 4500~~~        &  2070~~~          & 26.7  & 26.8     & 0.462     & 0.439          \\
\multicolumn{2}{ c }{$z_{850}$}  & \multicolumn{2}{c }{ 911}   & 9110~~~        &  9184~~~          & 27.7  & 27.7     & \multicolumn{2}{c }{0.540} \\
 $SZ$     &                        & \multicolumn{2}{c }{1070}   &21960~~~        &                   & 26.9  &          & 0.691     &                \\
\multicolumn{2}{ c }{$J$}          & \multicolumn{2}{c }{1259}   & 6480~~~        &  6480~~~          & 25.6  & 25.5     & \multicolumn{2}{c }{0.945} \\
\multicolumn{2}{ c }{$H$}          & \multicolumn{2}{c }{1656}   &13860~~~        & 12860~~~          & 24.7  & 24.7     & \multicolumn{2}{c }{1.412} \\
\multicolumn{2}{ c }{$K_{s}$}    & \multicolumn{2}{c }{2167}   &18990~~~        & 18990~~~          & 24.7  & 24.3     & \multicolumn{2}{c }{1.871} \\
\multicolumn{2}{ c }{3.6}        & \multicolumn{2}{c }{3577}   & 2400~~~        &  2400~~~          & \multicolumn{2}{c}{$\sim 0.2 \mu$Jy} & \multicolumn{2}{c }{2.790} \\
\multicolumn{2}{ c }{4.5}        & \multicolumn{2}{c }{4530}   & 2400~~~        &  2400~~~          & \multicolumn{2}{c}{$\sim 0.3 \mu$Jy} & \multicolumn{2}{c }{3.249} \\
\multicolumn{2}{ c }{5.8}        & \multicolumn{2}{c }{5788}   & 3600~~~        &  2400~~~          & \multicolumn{2}{c}{$\sim 1.2 \mu$Jy} & \multicolumn{2}{c }{3.737} \\
\multicolumn{2}{ c }{8.0}        & \multicolumn{2}{c }{8045}   & 3600~~~        &  2400~~~          & \multicolumn{2}{c}{$\sim 1.5 \mu$Jy} & \multicolumn{2}{c }{4.392} \\
\multicolumn{2}{ c }{24.0}       & \multicolumn{2}{c }{23680}  & 2700$^\star$~~ &  2700$^\star$~~   & \multicolumn{2}{c}{$\sim 10.0 \mu$Jy}  & \multicolumn{2}{c }{~} \\
\hline 
\label{tab_data}
\end{tabular}

\end{table*}

\subsection{ACS data}
New $z$-band (F850LP, denoted $z_{850}$ hereafter) imaging was obtained with the 
ACS camera onboard HST in  November 2004 (AC114) and July 2005 (Abell 1835).
The total observing time for AC114 and Abell 1835 were 9184 and 9110
seconds respectively.
While the AC114 field is centred at the central cluster galaxy, Abell 
1835 was observed off centre in order to avoid bright sources north of the
cluster. 
For the reduction of both data sets we used the IRAF/Pyraf
package {\it multidrizzle} on post-calibrated data \footnote{for
  details see http://stsdas.stsci.edu/pydrizzle/multidrizzle}.\\
For the calculation of the 1$\sigma$ detection limit, we applied the
same method as for all other bands \citep[see][]{r06}: measuring the corresponding 
standard deviation of the flux in randomly distributed
circular apertures of 1.5 arcsec diameter (approx. 3000). None of the aperture
positions lies within 3 arcsec to its closest neighbour or
sources which have at least a 3$\sigma$ detection. 

\subsection{IRAC \& MIPS data}
The 3.6, 4.5, 5.8, and 8.0 $\mu$m images were obtained using the
Infrared Array Camera \citep[IRAC;][]{Fazio2004}
while the 24 $\mu$m
images were obtained using the Multi-band Imaging Photometer for {\em
Spitzer} (MIPS; Rieke et al.\ 2004), both on board the {\em Spitzer}
Space Telescope \citep{Werner2004}.  The instruments, data, and
reduction procedures are described in \cite{Egami2006}.

\subsection{Chandra}
Both Abell 1835 and AC114 have been observed previously by {\it Chandra}.
AC114 has been observed once for a total exposure of 75 ksec (OBSID 1562).
A comprehensive analysis of the cluster X-ray properties based on this
dataset has been published previously in \cite{defilippis2004}.
In the case of Abell 1835, two short archival observations are available 
(OBSIDs 495 and 496) as well as a deep 200 ksec GO observation
(OBSIDs 6880, 6881, and 7370) obtained as part of this program.
All available data were reprocessed using CIAO 3.2 and the latest
calibration files available in CALDB 3.0. Standard screening was
applied to all event files to remove bad grades, bad pixels, and
background flares. After standard cleaning, the resulting net
exposures were 73 and 224 ksec for AC114 and Abell 1835, respectively.

Bright X-ray point sources in the fields of AC114 and Abell 1835 were
identified using the CIAO tool {\tt wavdetect}. No bright X-ray point
sources were detected within 5 arcsec of the ERO source positions.
To determine flux upper limits, source spectra were extracted in a 
2 arcsec radius aperture around each ERO source position. This aperture
captures virtually all of the Chandra PSF over the range of relevant 
off-axis angles. The local background for each source was determined
using an annular aperture from 2-4 arcsec surrounding the source
aperture. Count-weighted detector response (RMFs) and effective area
(ARFs) files were created for each extraction region using the CIAO
tools {\tt mkacisrmf} and {\tt mkwarf}, including the temporal,
spectral, and spatial dependence of the ACIS filter contaminant. 
For Abel 1835, source spectra and matching background, RMF, and ARF files
were produced for each OBSID separately and then fit jointly during
the spectral analysis. This analysis is discussed in more detail in 
Section 3.3. All spectral analysis was done using the ISIS \cite{houck2000} 
spectral fitting package and the XSPEC model library.

In the case of Abell 1835, spectra from all 5 observations (2 archival  datasets
and 3 datasets part of our program) were extracted for each source and
fit simultaneously. Each dataset was individually reprocessed using  CIAO 3.2
and calibration files available in CALDB 3.0.

\begin{figure*}[t!]
\begin{tabular}{lrl}
ID  & $^{a}$  &~~~~~~~~V~~~~~~~~~F702W~~~~~F814W~~~~~~~~$z_{850}$~~~~~~~~~~~~SZ~~~~~~~~~~~~~J~~~~~~~~~~~~~~~H~~~~~~~~~~~~~~K~~~~~~~~~3.6$\mu m$~~~~~~~~4.5$\mu m$~~~~~~~5.8$\mu m$~~~~~~~8.0$\mu m$\\
305 &  (1)& \includegraphics[width=16.0cm,clip]{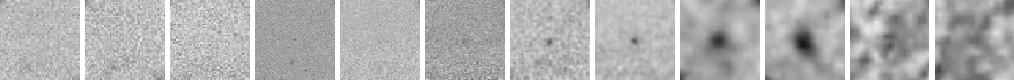} \\
319 &     & \includegraphics[width=16.0cm,clip]{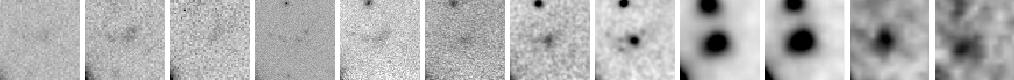} \\
347 &     & \includegraphics[width=16.0cm,clip]{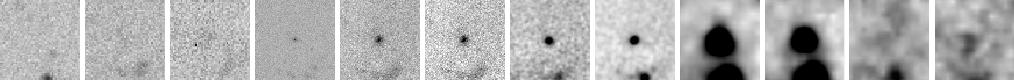} \\
532 &     & \includegraphics[width=16.0cm,clip]{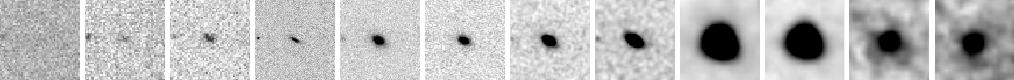}\\
676 &     & \includegraphics[width=16.0cm,clip]{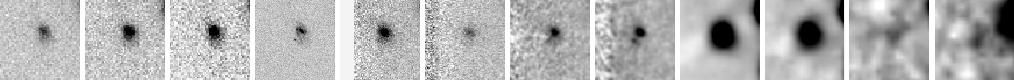}\\
1093&  (2)& \includegraphics[width=16.0cm,clip]{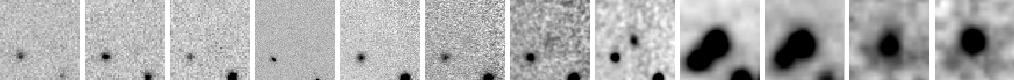}\\
    &     &   \\
    &     &   \\
\multicolumn{3}{l}{EROs listed in \citep{r06} and \citep{s07} based on 
  a 1$\sigma$ threshold in the used R-band}\\
311 & (17)& \includegraphics[width=16.0cm,clip]{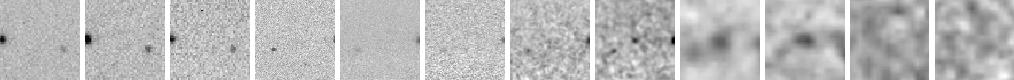} \\
314 & (11)& \includegraphics[width=16.0cm,clip]{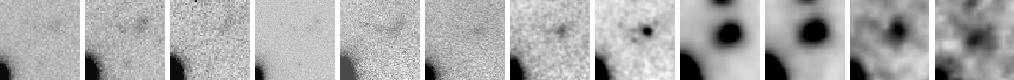} \\
454 & (10)& \includegraphics[width=16.0cm,clip]{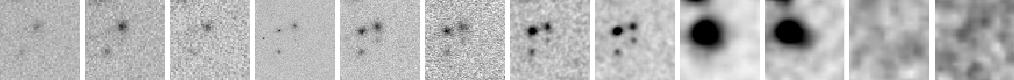} \\
493 &  (3)& \includegraphics[width=16.0cm,clip]{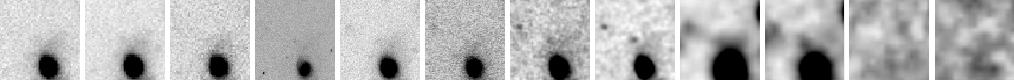}\\
504 &  (4)& \includegraphics[width=16.0cm,clip]{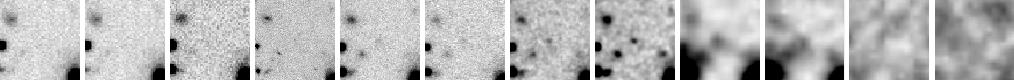}\\
\end{tabular}
\caption{Close-up images of EROs found in Abell 1835. Each of the panels is 20 arcsec across, North is up and East is left. Source
  number 1093 is the NIR-counterpart of the sub-mm source
  SMMJ14009+0252. $^{a}$ The numbers in brackets refer to \citet{r06}
  and \citet{s07}.}
\label{DLA}
\end{figure*}

\begin{figure*}[t!]
\begin{tabular}{rrl}
ID   & $^{a}$ &~~~~~~~~V~~~~~~~~~~~~~R~~~~~~~~~~~~~~~~~I~~~~~~~~~~~~~~~~~~$z_{850}$~~~~~~~~~~~~~~J~~~~~~~~~~~~~~~~H~~~~~~~~~~~~~~K~~~~~~~~~~~~~~3.6$\mu m$~~~~~~~~4.5$\mu m$~~~~~~~~5.8$\mu m$~~~~~~~~8.0$\mu m$\\
512  &      & \includegraphics[width=16.0cm,clip]{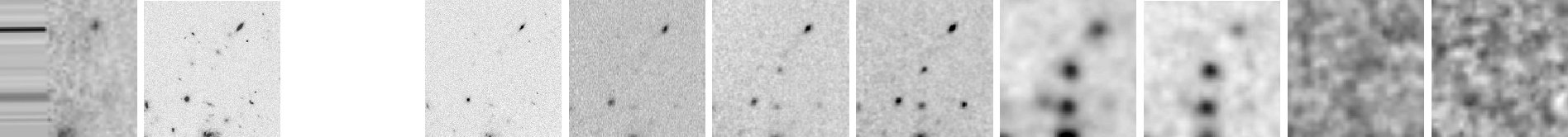} \\
572  &      & \includegraphics[width=16.0cm,clip]{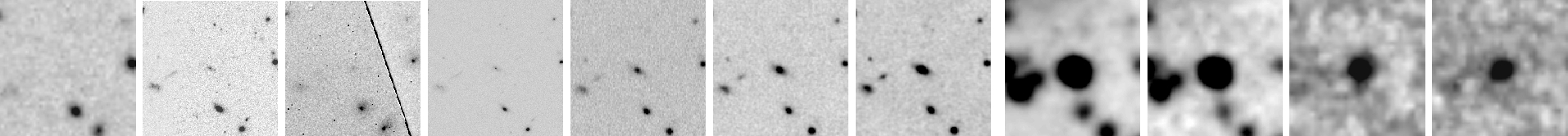} \\
632  &      & \includegraphics[width=16.0cm,clip]{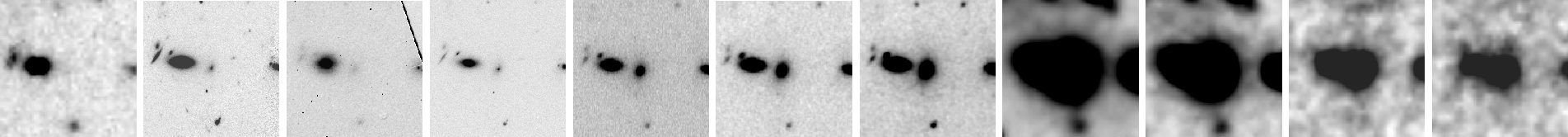} \\
680  &      & \includegraphics[width=16.0cm,clip]{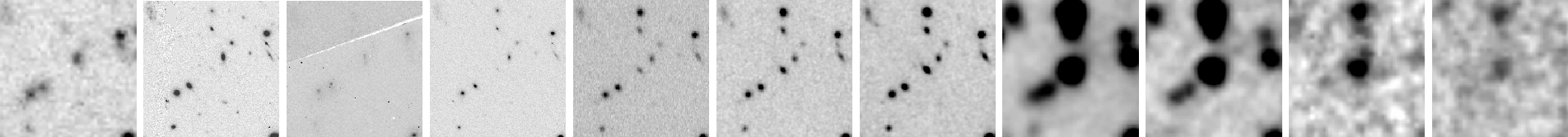} \\
707  &      & \includegraphics[width=16.0cm,clip]{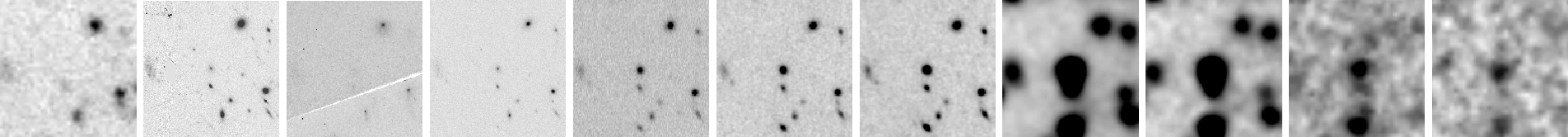} \\
862  &      & \includegraphics[width=16.0cm,clip]{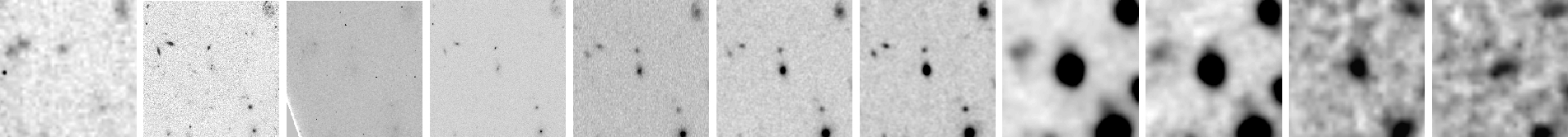} \\
1006 &      & \includegraphics[width=16.0cm,clip]{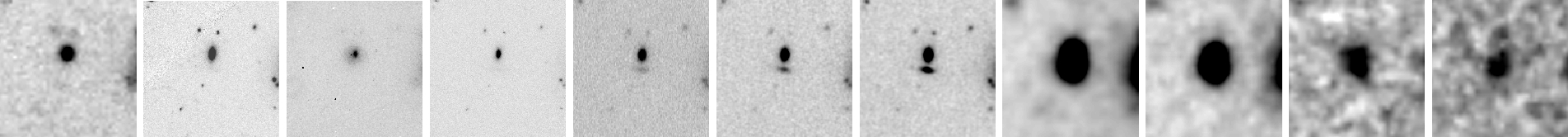} \\
1087 &      & \includegraphics[width=16.0cm,clip]{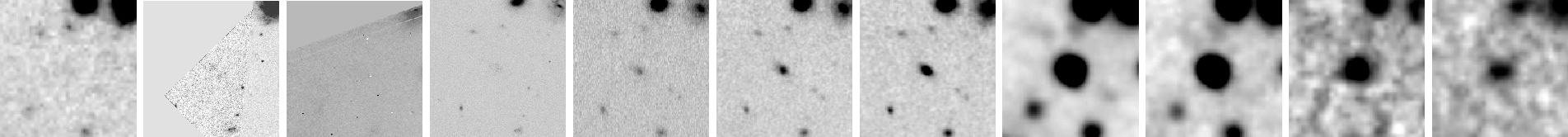} \\
1167 &   (1)& \includegraphics[width=16.0cm,clip]{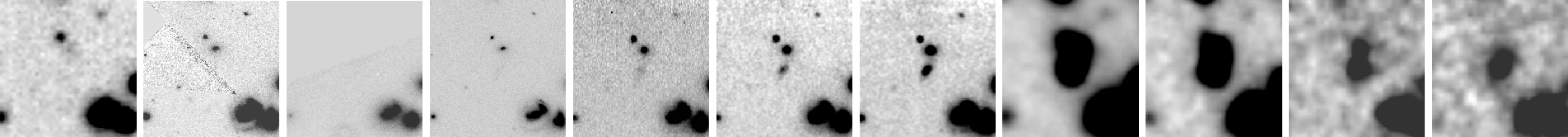} \\
\\
\end{tabular}
\caption{Same as Fig.\ref{DLA} for the EROs in AC114.
Source 512 lies outside the recorded I -band image.}
\label{DLA2}
\end{figure*} 

\section{Photometry}
\label{phot}
\subsection{Optical and near-infrared}
As our objective is to study EROs we proceed by identifying sources
in the ISAAC $Ks$-band image using
SE{\footnotesize XTRACTOR2.2.2} \citep{bertin96}\footnote{This
  software is freely available from http://terapix.iap.fr/} 
and requiring a 3$\sigma$ threshold above the background in at least 4 pixels. 
Note that our earlier work on these clusters, including the identification
of optical drop-out EROs, was done on an $H$-band selected sample \citep{r06}.
In contrast to \citet{r06} we use AUTO\_MAG instead of aperture photometry, 
mainly because some of our
EROs are quite extended and using large enough apertures might induce
additional problems due to the close proximity of other sources. As a
consequence we used the error provided by SE{\footnotesize XTRACTOR}
and not one based on the S/N characteristics in a fixed aperture
\citep[see][]{r06}. However, comparing the photometric errors for the
EROs already described in \citet{r06}, we found no significant
difference between the photometric error based on aperture photometry
or AUTO\_MAG. For the SED fitting a minimum photometric error of 0.1 mag
was generally used.\\
For the astrometry we used standard stars from the ESO-USNO-A2.0
catalogue to obtain correct coordinates (J2000). All coordinates are based on their
position in the $Ks$-band image. \\
The photometry in all ISAAC images ($SZJHK$) was done in the double-image mode of 
SE{\footnotesize XTRACTOR2.2.2}, using the $Ks$-band image as reference.\\
For objects which were not detected in a specific band, we substituted the detection
limit  for a 3$\sigma$ detection threshold as apparent magnitude.
Using our ERO criterion of $(R-Ks)\ge5.6$ we automatically compiled a
catalogue of ERO candidates. The $RJHK$ image of each of these candidates was then examined by 
eye in order to reject spurious detections, e.g. at the edge of the image or
candidates blended with another source.
This procedure resulted in the identification of 6 EROs in Abell 1835
and 9 in AC114. In \citet{r06} we found 7 and 1 resolved objects with extremely red optical to 
near-infrared colours in Abell 1835 and AC114 respectively, which qualified
as ERO. Five of the objects in Abell 1835 cannot be classified as ERO 
if we use the 3$\sigma$ detection limit for R-band non-detections. Dismissing the optical drop-out criteria applied in that paper and 
using a different photometry (see above) we now
find 4 additional EROs in Abell 1835 and 8 in AC114.
The postage stamps of our objects are shown in Figs. \ref{DLA} and
\ref{DLA2}. 

As result of the higher spatial resolution of ACS, A1835-\#319 
appears resolved into two sources.
It remains to be seen, whether or not we are looking at physically distinct
sources, or objects which are gravitational bound. For the adopted cosmology the
projected distance of 0.213 arcsec of the two components of \#319 corresponds
to 1.7 kpc, assuming a source redshift of 2.5 as indicated from their SED analysis
\citep{s07}. 
These estimates are based on the angular
distance without correction for lensing and hence state an upper limit 
 for the distance between the two components.\\

\renewcommand{\arraystretch}{1.5}
\begin{landscape}
\begin{table}[tb]
\caption{Optical and NIR photometry of EROs in Abell 1835. The astrometric position 
is given in J2000 coordinates.
Non-detections (lower limits) are 3 $\sigma$ values in all bands. The
number in brackets behind the ID number refers to the ID in
\citet{r06}. The magnitude errors given here are solely based
on the Sextractor output and are likely underestimated (cf.\ text).
Errors of 0.00 occur due to rounding effects. The last two columns
indicates whether the object satisfies the DRG criterion $J - K > 2.3$
or can be classified as IERO as described by \citet{yan04}
($f_{\nu}(3.6\mu m)/f_{\nu}(z_{850}) >$ 20).
ND stand for no data available, possibly due to blending.}
\begin{tabular}{rrccccccccccrr}
\hline 
\hline

 ID    &   &      RA       &    DEC       &  $V$      &   $R$     &    $I$    & $z_{850}$ &  $SZ$      &  $J$       & $H$       & $Ks$       &  DRG   &  IERO \\
       &   &               &              & 3$\sigma$& 3$\sigma$& 3$\sigma$ & 3$\sigma$ & 3$\sigma$ & 3$\sigma$ & 3$\sigma$ & 3$\sigma$ &        &       \\
\hline
 305        & ( 1)& 14:00:58.264 &  2:50:27.15 &  $>$26.9 &  $>$26.6  & $>$25.5  &  25.70$\pm$0.07 &  24.44$\pm$0.27 & 22.76$\pm$0.16 & 22.40$\pm$0.08 & 20.74$\pm$0.02  & no  &  yes\\  
 319$^{a}$ &     & 14:01:06.059 &  2:50:29.53 &  $>$26.9 &  $>$26.6  & $>$25.5  &  25.67$\pm$0.10 &  22.32$\pm$0.07 & 21.53$\pm$0.09 & 20.64$\pm$0.03 & 19.31$\pm$0.01  & no  &  yes\\  
 319$^{a}$ &     & 14:01:06.059 &  2:50:29.53 &  $>$26.9 &  $>$26.6  & $>$25.5  &  24.72$\pm$0.08 &  22.32$\pm$0.07 & 21.53$\pm$0.09 & 20.64$\pm$0.03 & 19.31$\pm$0.01  & no  &  yes\\  
 347        &     & 14:01:06.143 &  2:50:34.24 &  $>$26.9 & 26.59$\pm$0.37 & 23.38$\pm$0.18& 23.45$\pm$0.02 & 22.52$\pm$0.04 & 21.61$\pm$0.05 & 20.79$\pm$0.02 & 19.67$\pm$0.01  & no  &  no\\
 532        &     & 14:01:02.457 &  2:51:11.36 &  $>$26.9 & 24.95$\pm$0.12 & 22.57$\pm$0.06& 22.12$\pm$0.01 & 20.82$\pm$0.01 & 19.98$\pm$0.01 & 18.95$\pm$0.00 & 17.97$\pm$0.00  & no  &  yes\\  
 676        &     & 14:01:07.781 &  2:51:35.51 &  $>$26.9 &  $>$26.6  & $>$25.5  &  23.64$\pm$0.05 &  22.20$\pm$0.05 & 21.99$\pm$0.12 & 21.08$\pm$0.04 & 19.94$\pm$0.01  & no  &   ND\\\  
1093        & (~2)& 14:00:57.530 &  2:52:49.34 &  $>$26.9 &  $>$26.6  & $>$25.5  &   $>$26.5   &  24.08$\pm$0.26 &  $>$24.4   & 21.78$\pm$0.06 & 20.45$\pm$0.02  & yes  &  yes\\  
\multicolumn{14}{c}{~~~~}\\
\multicolumn{14}{c}{~~~~}\\
\multicolumn{14}{l}{EROs listed in \citep{r06} and \citep{s07} based  on a 1$\sigma$ detection threshold in the used R-band}\\
 311        & (17)& 14:01:04.995 &  2:50:27.73 &  $>$26.9 &  $>$26.6  & $>$25.5  &   $>$26.5   &   $>$25.7  & $>$24.4  & 23.51$\pm$0.16 & 22.11$\pm$0.03  & yes  &  yes\\  
 314        & (11)& 14:01:06.163 &  2:50:28.38 &  $>$26.9 &  $>$26.6  & $>$25.5  &   $>$26.5  &   $>$25.7  & 23.92$\pm$0.36 & 23.49$\pm$0.18 & 21.29$\pm$0.03  & yes  & ND\\  
 454        & (10)& 14:00:59.876 &  2:50:57.90 &  $>$26.9 &  $>$26.6  & $>$25.5 & 25.56$\pm$0.11 &  24.00$\pm$0.12 & 23.72$\pm$0.26 & 23.36$\pm$0.13 & 21.67$\pm$0.03  & no  &   ND\\ 
 493$^{a}$ & (~3)& 14:01:01.470 &  2:51:03.93 &  $>$26.9 &  $>$26.6  & $>$25.5 &  24.06$\pm$0.07 &  23.78$\pm$0.10 & 24.32$\pm$0.47 & 22.55$\pm$0.07 & 21.58$\pm$0.03  & yes  &   ND\\  
 493$^{a}$ & (~3)& 14:01:01.470 &  2:51:03.93 &  $>$26.9 &  $>$26.6  & $>$25.5 &  25.31$\pm$0.13 &  23.78$\pm$0.10 & 24.32$\pm$0.47 & 22.55$\pm$0.07 & 21.58$\pm$0.03  & yes  &   ND\\  
 504        & (~4)& 14:01:01.719 &  2:51:05.56 &  $>$26.9 &  $>$26.6  & $>$25.5 &  25.48$\pm$0.14 &  24.44$\pm$0.15 & 23.56$\pm$0.18 & 22.90$\pm$0.07 & 21.95$\pm$0.03  & no  &  no\\ 
\label{taba1835}
\end{tabular}
\end{table}
$^{a}$ Objects may appear twice because of multiple ACS
detections (\#319 and \#493).
\end{landscape}
\renewcommand{\arraystretch}{1.5}
\begin{landscape}
\begin{table}[tb]
\caption{Same as Table \protect\ref{taba1835} for the EROs in AC114.}

\begin{tabular}{rrcccccccccrr}
\hline 
\hline
 ID   &    &     RA      &       DEC     &     $V$   &     $R$   &     $I$   &  $z_{850}$&     $J$    &     $H$    &   $Ks$    &     DRG         &  IERO    \\
      &    &             &               & 3$\sigma$& 3$\sigma$& 3$\sigma$ & 3$\sigma$ & 3$\sigma$ & 3$\sigma$ & 3$\sigma$ &        &       \\
\hline
  512 &    &    22:58:53.278 &  -34:49:02.48 &  $>$27.3  & 26.49$\pm$0.12  &   $^{c}$      &  26.31$\pm$0.11 &  23.42$\pm$0.15 &  21.76$\pm$0.03 &  20.88$\pm$ 0.02  & yes  &  yes\\ 
  572 &    &    22:58:45.761 &  -34:48:47.89 &  $>$27.3  & 24.43$\pm$0.02  & 24.06$\pm$0.12 &  22.51$\pm$0.02 &  20.56$\pm$0.02 &  19.42$\pm$0.01 &  18.51$\pm$ 0.00  & no\  &  yes\\
  632 &    &    22:58:46.664 &  -34:48:31.89 &  $>$27.3  & 23.04$\pm$0.01  & 21.55$\pm$0.03 &  20.57$\pm$0.01 &  18.64$\pm$0.01 &  17.64$\pm$0.00 &  16.75$\pm$ 0.00  & no  &  no\\
  680 &    &    22:58:51.361 &  -34:48:26.90 &  $>$27.3  & 26.65$\pm$0.13  & $>$25.6    &  24.51$\pm$0.07 &  21.90$\pm$0.05 &  20.30$\pm$0.01 &  19.34$\pm$ 0.01  & yes  &  yes\\
  707 &    &    22:58:51.357 &  -34:48:18.55 &  $>$27.3  & 24.27$\pm$0.03  & 24.17$\pm$0.15 &  22.36$\pm$0.02 &  20.52$\pm$0.02 &  19.50$\pm$0.01 &  18.60$\pm$ 0.00  & no  &  no\\ 
  862 &    &    22:58:52.560 &  -34:47:56.54 &  $>$27.3  & 25.18$\pm$0.04  & 23.64$\pm$0.09 &  22.66$\pm$0.02 &  20.83$\pm$0.02 &  19.72$\pm$0.01 &  18.69$\pm$ 0.00  & no  &  yes\\ 
 1006 &    &    22:58:49.014 &  -34:47:26.53 &  $>$27.3  & $>$27.7         & $>$25.6    &  24.27$\pm$0.04 &  21.19$\pm$0.04 &  20.00$\pm$0.01 &  19.01$\pm$ 0.01  & yes  &  ND\\ 
 1087 &    &    22:58:51.727 &  -34:47:07.85 &  $>$27.3  & $>$24.9$^a$
 & 25.5$\pm$0.33 &  23.77$\pm$0.04 &  20.92$\pm$0.03 &  19.63$\pm$0.01
 &  18.66$\pm$ 0.00  & yes  &  yes\\ 
 1167 & (1)&    22:58:49.775 &  -34:46:55.00 &  $>$27.3  & $>$24.9$^b$
 & $>$25.6    &  24.55$\pm$0.07 &  21.26$\pm$0.04 &  19.75$\pm$0.01 &
 18.62$\pm$ 0.00 &   yes   & ~yes
\label{tabac114}
\end{tabular}
\end{table}
~\\
$^a$ Since the effective exposure time is lower in this region a lower detection limit
has been applied (cf.\ Section \ref{aperture}).\\
$^b$ Object located where two regions with different exposure times meet.
The detection limit for the shorter exposure has been adopted (see 
Section \ref{aperture} for more details).\\
$^c$ Object lies outside of image
\end{landscape}
\renewcommand{\arraystretch}{1.0}

\subsection{IRAC and MIPS photometry}

The IRAC photometry used a circular aperture with a radius of 2\farcs4
with a sky background annulus of 2\farcs4--7\farcs2 in radius.  The
point-source aperture corrections were applied, which were 1.213,
1.234, 1.379, and 1.584 at 3.6, 4.5, 5.8, and 8.0 $\mu$m,
respectively, based on the IRAC Data Handbook.

The MIPS 24 $\mu$m photometry used a circular aperture with a radius
of 6\arcsec\ with a sky background annulus of 6\arcsec--13\arcsec\ in
radius.  The corresponding point-source aperture correction was 1.698
based on the MIPS instrument Web site.  
For several EROs we encountered severe blending problems (see
Tab.\ref{tab1}) and hence these were not included in our discussion.

\begin{table*}
\caption{
IRAC and MIPS fluxes for EROS found in Abell 1835 and AC114. 
$^\star$ The source positions at 5.8 and 8 $\mu$m are slightly displaced
    from those at shorter wavelengths.  Therefore, there is a
    possibility that the emission at $> 4.5 \mu$m is not related to
    the ERO but to the fainter sources to the east. Objects for
      which we encountered blending problems are not included in
      Fig. \ref{CMD1} and \ref{IK45}.}
\begin{tabular}{rrrrrrr}
\hline 
\hline
ID       &    & 3.6$\mu$m     &   4.5$\mu$m     &   5.8$\mu$m    &    8.0$\mu$m    &     24$\mu$m  \\
Abell 1835    &     &   $\mu$Jy       &   $\mu$Jy       &   $\mu$Jy      &    $\mu$Jy      &     $\mu$Jy   \\   
\hline
305      & (1)&   2.9$\pm$ 0.2  &   4.0$\pm$ 0.2  &    $<$ 3.6     &      $<$4.5     &  $<$  30.    \\             
319      &    &  19.7$\pm$ 0.4  &  22.5$\pm$ 0.4  &  26.5$\pm$ 1.7 &  18.1$\pm$ 1.6  &  272$\pm$9$^\star$ \\
347      &    &  15.7$\pm$ 0.3  &  13.9$\pm$ 0.3  &    $<$ 3.6     &   8.0$\pm$ 1.5  &  $<$  30.    \\ 
532      &    &  73.6$\pm$ 0.3  &  75.3$\pm$ 0.3  &  51.3$\pm$ 1.5 &  37.1$\pm$ 1.5  &  $<$  30.   \\ 
676      &    &      \multicolumn{4}{c}{blended}                                     &             \\
1093     & (2)&  14.4$\pm$ 0.2  &  23.0$\pm$ 0.3  &  37.6$\pm$ 1.5 &  50.9$\pm$ 1.6  &  320$\pm$11   \\
         &                 &                 &                &               &                    \\                     
311      &(17)&   2.4$\pm$ 0.2  &   1.6$\pm$ 0.2  &    $<$ 3.6     &                 &  $<$  30.    \\
314      &(11)&   \multicolumn{4}{c}{blended}                                        &             \\                       
454      &(10)&  \multicolumn{4}{c}{blended}                                         &             \\
493      & (3)&  \multicolumn{4}{c}{blended}                                         &             \\
504      & (4)&   1.9$\pm$ 0.2  &   1.0$\pm$ 0.3  &    $<$ 3.6     &                 &  $<$  30.   \\          
         &    &                 &            &                     & 
              \\
\hline 
\hline
AC114    &    &                 &            &                     &
\\   
\hline 
 512     &    &   5.2$\pm$0.2  &     5.3$\pm$0.2 &   7.1$\pm$1.2  &  $<$4.5         &   $<$30.   \\        
 572     &    &  56.0$\pm$0.3  &    57.4$\pm$0.3 &  51.0$\pm$1.3  &  43.8$\pm$1.4   & 124.9$\pm$7.9   \\        
 632     &    & 210.8$\pm$1.0  &   194.6$\pm$0.9 & 131.2$\pm$1.7  &  92.1$\pm$2.2   &   $<$30.   \\           
 680     &    &  27.0$\pm$0.2  &    29.9$\pm$0.3 &  31.8$\pm$1.1  &   16.9$\pm$1.4   &   $<$30.   \\        
 707     &    &  46.2$\pm$0.3  &    43.8$\pm$0.3 &  25.1$\pm$1.2  &  23.2$\pm$1.4   &   $<$30.   \\        
 862     &    &  39.0$\pm$0.2  &    35.2$\pm$0.3 &  28.7$\pm$1.1  &  22.5$\pm$1.4   &  58.0$\pm$10.4 \\
1006     &    & \multicolumn{2}{c}{blended}      &  36.0$\pm$1.2  &  27.5$\pm$1.4   &  51.0$\pm$7.9    \\            
1087     &    &  48.2$\pm$0.2  &    55.3$\pm$0.3 &  44.0$\pm$1.2  &  36.5$\pm$1.5   & 183.7$\pm$7.4   \\        
1167     & (1)&  67.3$\pm$0.5  &    64.4$\pm$0.5 &  50.3$\pm$1.5  &  44.9$\pm$2.2   & 189.0$\pm$8.9   \\            
         &    &                 &                 &                &                 &         \\       
\hline 
\label{tab1}
\end{tabular}
\end{table*}

\subsection{Chandra photometry}
\label{chandra}
None of the ERO sources were detected in the fields of AC114 and A1835.
In order to place upper limits on the possible X-ray emission from the
EROs, we have compared the extracted spectra discussed in Section 2.3 at each ERO source position with a number of spectral models for the assumed
underlying spectral energy distribution. These flux distributions were fit
to a simple power law model including foreground Galactic absorption. 
Such a model would be expected if the intrinsic X-ray spectra of the
ERO was dominated by AGN emission. The absorbing column was fixed to
the Galactic value for each cluster and held fixed during the fitting
procedure. Due to the low number of counts associated with a given
source, the spectral index of the power-law model was similarly held
fixed during the fitting. Values of 1.0, 1.4, and 2.0 were considered
for the photon spectral index and different energy ranges were
considered for comparison with other data from the literature.  
The resulting flux limits for each ERO source (at a 3-sigma
  level) are listed in Table \ref{tab_chandra}.

\begin{table}[htb]
\caption{X-ray flux limits for the EROs in Abell 1835 and AC114 computed 
for two different energy bands and different values of the photon energy index
$\Gamma$. All upper limits are in units of ergs
s$^{-1}$cm$^{-2}$ and refer to a 3$\sigma$ detection. }
\begin{tabular}{rccc}
\hline 
\hline 
 ID       & \multicolumn{2}{c}{(0.5-7.0 keV)}   & (2.0-10.0 keV)\\
Abell 1835    &  $\Gamma$=1.0  &   $\Gamma$=2.0  &  $\Gamma$=1.4   \\
\hline 
 305    &     3.14e-16  &   2.21e-16  &   2.64e-16  \\
 319    &     5.44e-16  &   3.85e-16   &  4.60e-16  \\
 347    &     9.98e-16  &   7.08e-16   &  8.43e-16  \\
 532    &     3.81e-16  &   2.70e-16  &   3.21e-16  \\
 676    &     1.71e-15  &   1.22e-15  &   1.45e-15  \\
1093    &     5.47e-16  &   3.87e-16  &   4.63e-16  \\
        &            &             &            \\
 311    &     5.42e-16  &   3.85e-16   &  4.60e-16  \\
 314    &     3.72e-16  &   2.64e-16   &  3.16e-16  \\
 454    &     3.98e-16  &   2.81e-16   &  3.36e-16  \\
 493    &     4.72e-16  &   3.34e-16   &  3.99e-16  \\
 504    &     5.11e-16  &   3.61e-16   &  4.30e-16  \\
        &            &             &            \\
        &            &             &            \\
\hline
\hline                                     
AC114   &            &             &            \\
\hline
 512    &     8.79e-16  &   4.52e-16   &  6.55e-16  \\
 572     &    2.37e-15  &   5.75e-16   &  1.26e-15  \\
 632     &    3.25e-15  &   9.74e-16   &  2.06e-15  \\
 680     &    4.32e-15  &   1.70e-15   &  3.05e-15  \\
 707     &    1.09e-15  &   5.58e-16   &  8.13e-16  \\
 862     &    3.05e-15  &   1.55e-15   &  2.26e-15  \\
1006     &    1.25e-15  &   6.31e-16   &  9.25e-16  \\
1087     &    9.72e-16  &   4.91e-16   &  7.19e-16  \\
1167     &    8.63e-16  &   4.38e-16   &  6.40e-16  \\
        &            &             &            \\
\hline 
\label{tab_chandra}
\end{tabular}
\end{table}

\section{Empirical properties of our EROs and comparisons with other samples}
\label{s_samples} 
As mentioned above our search for EROs with $R-Ks>5.6$ in Abell 1835 
and AC114 has yielded 15 (16) objects in total, depending if or not sources appearing
as double in the ACS images are counted. Two of the additional EROs
(\#347 \& \#532) detected in
Abell 1835 were not included in \citet{r06} and \citet{s07} due to
the additional selection criterion, optical non-detection, imposed in
these papers. Two more objects (\#319 \& \#676) were previously excluded during the by eye
examination (as described in section \ref{phot}). \\
Six of the additional EROs in AC114 can be detected in R using the initial
detection threshold of 1$\sigma$ and hence were not included in
\citet{r06} or \citet{s07}. The remaining two new sources were
previously excluded during the visual examination, either due to a
close by bright object (\#1006) or
because its position at the edge of the image (\#1087).\\
We now discuss the properties of the EROs and compare them to other
samples and to related objects. At this point we note that we do not differentiate between the various $K$-band filters. \\
\subsection{EROs with very red $R-Ks$ colours}
\label{aperture}
\begin{figure}[htb]
   \centering\includegraphics[width=8.8cm]{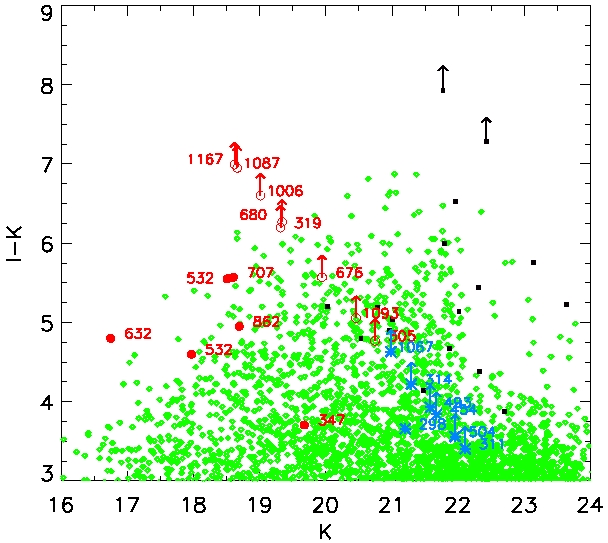}
   \caption{$I-K~vs~K$ colour-magnitude diagram for our sample of EROs
     (red), IEROs in the HUDF \citep[black,][]{yan04} and EROs found in
     GOODS-MUSIC \citep[green,][]{grazian07}.  Blue stars identify 
       the objects described in \citet{r06} and \citet{s07}. Arrows
     mark upper limits, open circles denote EROs without R-band detection.}
    \label{eros2}
   \end{figure}

Fig. \ref{eros1} and \ref{eros2} show some of the photometric
properties of our EROs in comparison to other samples.
These figures show that there are some sources, both in Abell 1835 and AC114, 
which have very red colours both with $(R-Ks) \ga 7$ and $(I-Ks)\ga 6$ 
(adopting $3\sigma$ limits for the non-detected bands). Such
relatively bright ($Ks < 20.3$) and red
sources were not found in other surveys, e.g. HUDF \citep{yan04}, MUNICS
\citep{longhetti05} and GOODS-MUSIC
\citep{grazian2006,grazian07} although their depth is sufficient to detect such red objects. However, \citet{sawicki2005} report 5 EROs with $R-K_{s}>$7.0 at similar magnitudes as our objects. The origin of these apparent
differences is not clear. However, it has to be recognised that all samples except GOODS-MUSIC are relatively small and lack statistical
significance.

\subsection{ERO classification (starburst vs.\ old population)}

\begin{figure}[htb]
   \centering\includegraphics[width=8.8cm]{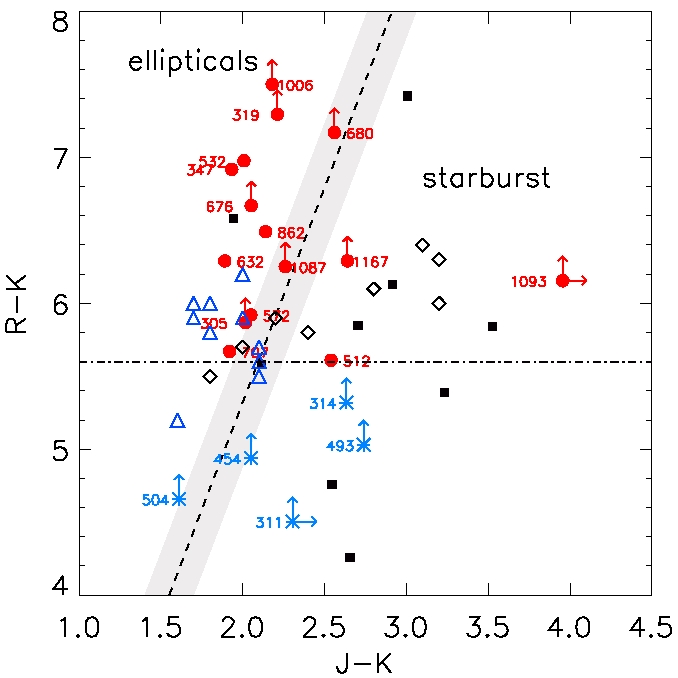}

   \caption{Colour-colour plot of EROs found in Abell 1835 and
     AC114. Blue stars indicate objects which were classified as ERO 
     based on a 3$\sigma$ non- detection \citep[see][]{r06,s07}. Black diamonds show EROs found by Takata et
     al. (\citeyear{takata03}) in the vicinity of SCUBA source SMM
     J0452-0301. Black squares mark IEROs in the HUDF
     \citep{yan04} and blue triangles show EROs found in the MUNICS
     survey \citep{longhetti05}.The dashed line shows the separation between old
     passive galaxies and dusty starburst, according to Pozzetti \&
     Mannucci (\citeyear{pozzetti00}). The shaded area
     represents the gap between the ERO populations ($\approx$0.3
     mag). Arrows indicate the $R-Ks$ or $J-Ks$ colour
     to be a lower limit due to the non-detection in $R$ or $J$ respectively 
     (see Tabs. \ref{taba1835} and \ref{tabac114}).}
    \label{eros1}
   \end{figure}

Although colours alone cannot provide the same strong constrains  on their nature and
photometric redshift as SEDs or spectra, we used the colour based
classification scheme introduced by \citet{pozzetti00} for a first classification of our
sample. This photometric method uses the $(R-K)$
{\it vs.} $(J-K)$ colour plane to separate
 between galaxies with old stellar populations and dusty starbursts, assuming 
a redshift range of $1 \la z \la 2$.
The corresponding colour-colour diagram of our EROs is shown in Fig. \ref{eros1}.
For comparison
we have also included EROs found by the
MUNICS survey \citep{longhetti05}, IEROs in the HUDF \citep{yan04} and a
sample of EROs found by \citet{takata03} in the
field of the submillimeter source SMM J04542-0301 (cluster
MS0451.6-0305). 

However, taking  other information into account this simple classification
scheme does not always yield consistent results.
For example, several objects classified as ``elliptical'' on the basis of 
Fig.\ \ref{eros1} are detected at 24 \micron\ -- incompatible with
an old and dust free population. Furthermore the SED analysis  (cf.\ Sect.\
\ref{s_fit}) of these objects and several other objects in the ``ellipticals'' 
region shows that they are more likely dusty bursting objects.
Nevertheless, one has to keep
in mind that many of our objects have colours close to the separation
line (as calculated by \citealt{pozzetti00}) and that the gap between
both populations is approximately 0.3 magnitudes wide (shaded area in Fig.\ref{eros1}). In this respect,
many of our less extreme EROs
could also be classified as starburst. Also, Pozzetti \& Mannucci
include exponential declining SFH up to decay times of $\tau \sim$ 0.3
Gyr in their models of evolved populations, while we only consider
instantaneous burst scenarios for evolved populations.\\
~\\
\subsection{Redshift estimates for EROs with IRAC and/or MIPS detection}
\label{EROSIRAC}
According to \citet{wilson04} an ERO selection equivalent to $R-K>5$ 
(less red than our colour threshold) using the IRAC 3.6$\mu$m band would require an $R-[3.6]$
colour redder than 6.6 (Vega) or 4.0 (AB magnitude). 
All our IRAC detected EROs fulfil this criterion (generally $R-[3.6]_{Vega}> 7.4)$.
The combination of $Ks$-band and IRAC bands also allow a rough estimate 
of redshift, based on the shift of the 1.6 $\mu m$ bump. This spectral 
feature can be found in the spectra of all galaxies with the exception of AGN
dominated SEDs, and hence it can be used to estimate the
photometric redshift \citep{sawicki02}.
In practice using colour criteria of $(K-[3.6])_{Vega}>0.9$ and $([3.6]-[4.5])_{Vega}<0.47$ 
limits the photometric redshift interval to $0.6<z<1.3$, while $(K-[3.6])_{Vega}>0.9$ 
and $([3.6]-[4.5])_{Vega}>0.47$ should select galaxies with redshifts above 1.3,
according to \citet{wilson04}.
The corresponding colour-colour plot showing our objects and comparison
samples is given in Fig.\ref{CMD1}. All objects except \#305 satisfy this $K-[3.6]$ criterion.
From their red $([3.6]-[4.5])$ we expect that \#1093 is clearly
above $z > 1.3$. In fact, detailed SED modelling yields photometric redshifts
estimates of $z_{\rm phot} \sim$ 2.0 and 2.8--3. respectively (see
Section \ref{s_fit} and \citet{s07}) in agreement with this simple
criterion. 
However, for the remaining objects with  $([3.6]-[4.5])$ close to the 
limit proposed by \citet{wilson04}, there is only partial agreement for 
the separation of sources above or below $z=1.3$ using the two methods,
as can be seen by comparison with Table \ref{t_props}.
More details on the photometric redshift determination of the EROs are
given in Sect.\ref{s_fit} and \citet{s07}.\\
From a study of 24-\micron\ selected objects \citet{magloocc2007} suggest 
that objects with extreme 24-\micron\ to
$R$-band ratios of $\log{F_{24}/F_R} \la$ --3 are likely $z \sim$ 1.6--3.
All 6 MIPS detected objects, except \# 572 fall in this category.
For \#1093 both the colour criteria discussed above and
SED modelling agree with this classification. For the 3 remaining
objects (\# 1006, \# 1087, \# 1167) a more complete SED fitting
yields, however, photometric redshifts between $\sim$ 0.9 and 1.5.
Such a simple criterion may thus overestimate the redshift of sources
with extreme IR/optical flux ratios.
However, since our objects have quite faint MIPS fluxes, below the levels
of $\sim$ 0.3 mJy discussed by \citet{magloocc2007} and \citet{houck2005} 
their criterion may be correct for more luminous sources.

\subsection{Comparison of EROs with other galaxy populations}

A significant overlap between different galaxy populations selected e.g.\
according to ERO, DRG, and other criteria is known to exist.
In our case, e.g.\ 6 of our 15 EROs satisfy also the DRG selection criterion
$J-K \ge 2.3$ as indicated in Tables \ref{taba1835} and
\ref{tabac114}, while 11 objects show a large 3.6\micron/$z_{850}$ flux
ratio used by \citet{yan04} to classify IRAC selected objects (IEROs).

   \begin{figure}[htb]
   \centering \includegraphics[width=8.8cm]{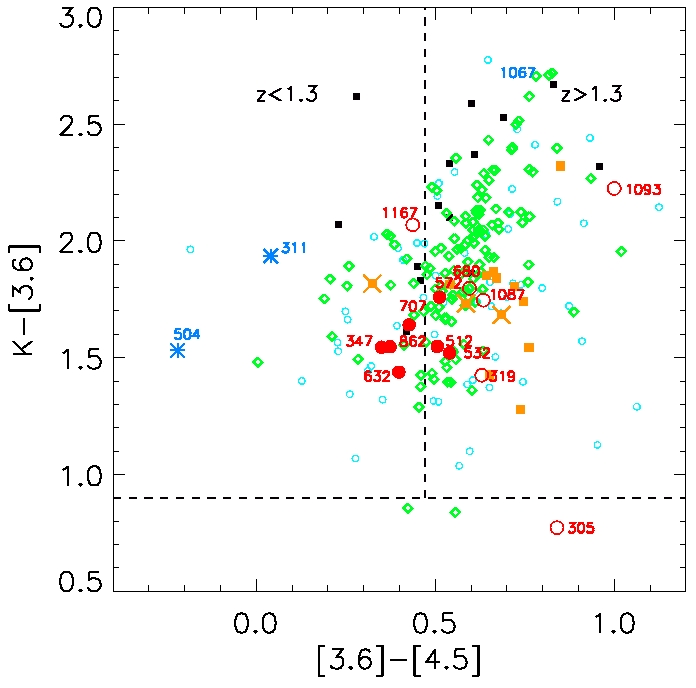}
   \caption{$K-[3.6] {\it vs.} [3.6]-[4.5]$ colour-colour diagram for
     extremely red galaxies. 
     The red symbols represent our work. Filled: EROs with R-band
     detection, open: no R-band detection. We also includes the ERos
     from \citep{r06,s07}(blue stars). 
 The blue circles denote EROs selected on the basis of
 their red ($R-[3.6])_{Vega}>$ 6.6 or ($K-[3.6])_{Vega}>$1.6) colour
 \citep{wilson04}. Green and orange symbols show distant red galaxies (DRG)
 with $J-K>$2.3  from \citet{papovich06} and \citet{labbe05}
 respectively. Black squares indicate IEROs by \citet{yan04}. }
   \label{CMD1}
   \end{figure}

\begin{figure}[htb]
   \centering \includegraphics[width=8.8cm]{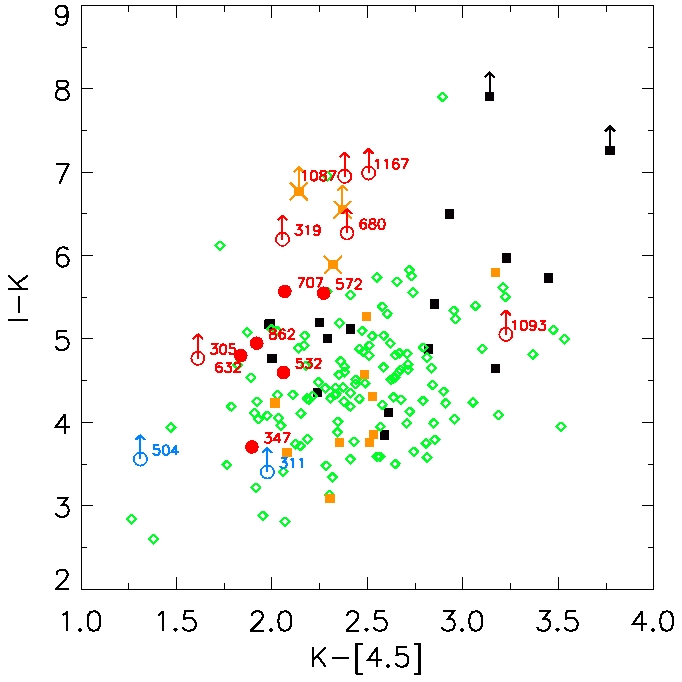}
   \caption{Observed $I-K vs. K-4.5\mu m$ colour-colour diagram. Equal 
     symbols to Fig. \ref{CMD1}. Arrows indicate I-band
     non-detections, open circles non-detections in R.}
   \label{IK45}
   \end{figure}

Considering e.g.\ the near-IR and IRAC colours shown in Fig.\ \ref{CMD1}, 
we do not find a distinct difference to 
other populations: EROs \citep{wilson04}, DRGs
\citep{papovich06} and IEROs \citep{yan04}. 
However, from this and from Fig.\ \ref{IK45} it is clear that all
sources without $R$-band detection lie at the outer regions in these
plots, indicating somewhat more extreme colours than DRGs, which
are however shared by some IEROs. 
Overall most of them correspond to objects with very strong extinction
as obtained from the SED analysis in Section \ref{s_fit}.

In Fig. \ref{IK45} we show the $I-K$ versus $K-4.5\mu m$
colour, which Labb\'e et al. \citeyear{labbe05} use to separate
distant red galaxies (DRGs) from $z\sim$2.5 Lyman break galaxies. All
samples, DRGs  (\citet{labbe05,papovich06},  orange and green
symbols), IEROS (\citet{yan04}, black symbols) and our EROs (red) occupy the
same colour space, with EROs without $R$-band detection again occupying the
outer regions. We also include the ERO sample of \citet{wilson04}. Three of the DRGs by  Labb\'e et
al. (\citeyear{labbe05}) are thought to have old stellar populations
(orange stars), due to their very red $I-K$ colour. The 5 EROs in 
our sample, which have similar or even redder $I-K$ colours and comparable $K-[4.5]$
colours, were classified as ``old evolved'' galaxies due to their blue 
$J-K$ colour.
However, from SED fitting (see below) we find that the majority of them
are best fit with GRASIL spectral templates of very dusty star forming galaxies, 
which is also supported by their detection at 24 \micron.
Such extreme templates were not considered by \citet{labbe05}.
This shows that not all objects with such extreme $I-K$ colours
are ``old and dead'' galaxies, as suggested by \citet{labbe05}.
In \citet{s07} we have also shown that the bulk of the IEROs of \citet{yan04}
are more likely dusty starbursts than old composite stellar populations.
Detailed SED analysis including deep mid-IR observations may thus be 
needed to determine accurately the fraction of ``old and dead'' galaxies
among red distant galaxies, as also pointed out by \citet{kriek2006}.

  \begin{figure}[htb]
   \centering\includegraphics[width=8.8cm]{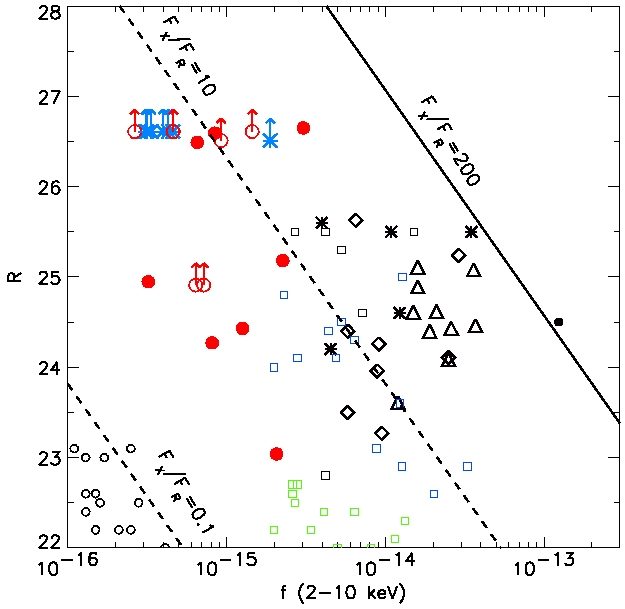}
   \caption {$R$ magnitude vs. 2-10 keV flux (in units of erg
     cm$^{-2}$s$^{-1}$) for our sample (red symbols, see Tab. \ref{tab_chandra}), and for other X-ray 
     emitting EROs. As none of our sources (red and blue symbols)
       was detected with Chandra, their Xray flux represents an upper
       limit. For a better clarity of the plot we have obmitted the
       arrows in the horizontal direction. Triangles (black): \citet{mignoli2004}, diamonds (black):
     \citet{brusa2005}, small circles (black): \citet{alexander2003}, stars:
     \citet{roche03}, large dot: XBS J0216-0435
     \citep{severgnini2006,della2004}, open squares:
     \citet{mainieri2002}. The additional objects from
       \citet{r06} and \citet{s07} are
       indicated with blue stars. 
     The two dashed lines define the region where unobscured type 1 AGNs typically lie
     \citep[see][]{fiore2003,maccacaro1988}.}
     \label{rxray}
   \end{figure}

  \begin{figure}[htb]
   \centering\includegraphics[width=8.8cm]{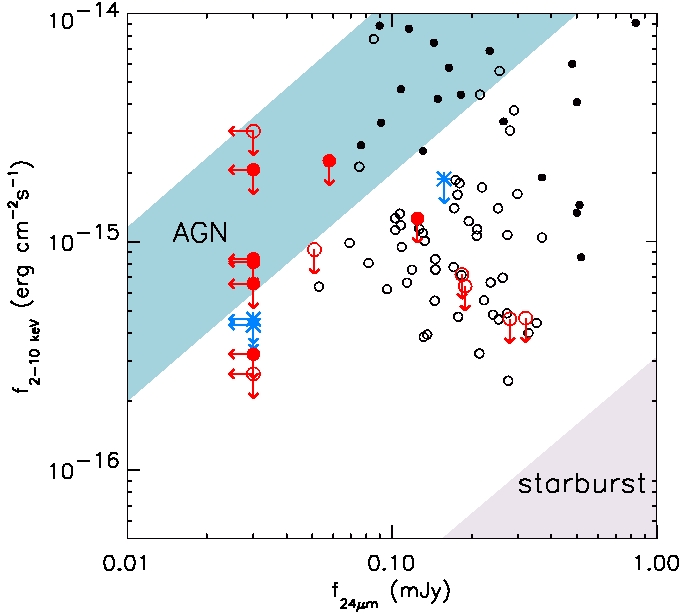}
   \caption {The 24$\mu$m flux vs. 2-10 keV X-ray flux (red: our work,
     filled: with $R$-band detection, open: no $R$-band
     detection). Arrows 
     assign upper limits, both in X-ray and MIPs flux. Black symbols show X-ray
     selected sources with 24$\mu$m counterparts in the CDF-S \citep{alonsoherr2006}
    . Open black circles show sources not
     detected in hard X-rays. The  Chandra 2-8 keV fluxes have
     been converted to 2-10 keV fluxes assuming a power-law with
     photon index $\Gamma$=1.4.
     The blue shaded area is the extrapolation of the median hard
     X-ray-to-mid-IR ratios of local (z$<$0.12) hard X-ray selected
     AGNs with mid-IR emission \citep{piccinotti1982}. The purple area
     is the extrapolation of local starburst galaxies from Ranalli et
     al. (\citeyear{ranalli2003}). The extrapolations were taken from
     Alonso-Herrero et al. (\citeyear{alonsoherr2004,alonsoherr2006}).}
   \label{xraymips}
   \end{figure}

\subsection{Starburst vs.\ AGN classification from X-rays}
\label{AGNclass}
None of our EROs is detected in X-rays above the background of the galaxy cluster emission. 
However, thanks to the depth of the observations and the location of the objects
away from the maximum cluster emission, the upper limits provide some
interesting information on the nature of the EROs.

In Fig. \ref{rxray} we plot the optical ($R$-band) flux or upper limits of
our objects versus their X-ray flux limits.  Also shown are other ERO samples
from the literature \citep{mignoli2004,brusa2005,severgnini2006}
and curves of constant X-ray to optical flux ratios; the range between 
$F_X/F_R=0.1$ and 10 is typical for unobscured type 1 AGN. 
The majority of our objects detected in $R$ are close to the border or outside
of this region, indicating that they are likely not unobscured AGN. However, deeper X-ray observations might position our sources in the part of the plot occupied by those objects. For the other objects
this comparison relying on the $R$-band is not a strong constraint on
their nature. More interesting is the comparison of the MIPS 24 $\mu$m flux with
the X-rays, which is very useful to compare AGN or starburst dominated
objects, as shown e.g. by \citet{alonsoherr2004}.
These data are shown in Fig. \ref{xraymips} together with the regions
of typical X-ray/mid-IR fluxes of local hard X-ray selected AGNs and
local starbursts taken from \citet{alonsoherr2004}.
Clearly the majority of 24 \micron\ detected sources , i.e.\ 5 out of 7, have 
X-ray limits excluding AGNs and compatible with expectations from 
local starbursts.
For the remaining objects we cannot conclude firmly on their nature.
However, no signature of an AGN is detected. In fact if some of these objects
turned out to be AGN they would correspond to very faint AGN, given
their reasonably well established redshift.
For our MIPS detected objects the relatively small 8/24 \micron\ flux 
ratio is also compatible with starburst dominated objects
\citep[cf.][]{magloocc2007}.
We conclude that the bulk of our EROs are more likely starburst than AGN
dominated at near- to mid-IR wavelengths.\\

\subsection{Magnification}
Our only selection criteria is based on $R-K$ colour which is not
influenced by the magnification effects. However, in order to compare
the number counts with other surveys, either in the field of lensing
clusters (e.g. \citealt{smith2002}) or large surveys
\citep{simpson2006,daddi2000,smail2002} one has to correct for the
  magnification of the source flux and the dilution of the source
  plane.  Magnification  maps were  derived following the
  procedure described in \citet{r06} using the mass
  models  of Abell 1835 (similar to Smith et al. 2005) and AC114
  \citep{natarajan1998,campusano2001}. Given the position of each
  object (in terms of RA and DEC), we then determined the
  magnification factors.  
  Table \ref{mag} lists the
necessary correction if the source plane lies at redshift
0.5, 1.0, 1.5, 2.0, 3.0 and 7.0
respectively. As the field of Abell 1835 is positioned away from the cluster
centre, most of the sources experience a relative small correction. 
These magnification factors $\mu$ also have to be taken into account
to compute absolute quantities, such as the stellar, SFR, etc.\
derived in Section \ref{s_fit}.

\begin{table*}[htb]
\caption{Magnification factors $\mu$ from the lensing models of  Abell 
  1835 
  and AC114 predicted for various source redshifts $z_{s}$. The values 
  of $\mu$ are dimensionless magnification factors, and not in
  magnitudes. The magnification factors derived for $z_{s}=1.5$ were
  used for the calculation of corrected number counts.}
\label{mag}
\begin{tabular}{rrcccccc}
\hline
\hline
 ID   &     & $z_{s}=0.5$  & $z_{s}=1.0$ & $z_{s}=1.5$ & $z_{s}=2.0$& $z_{s}=3.0$& $z_{s}=7.0$  \\
\hline
\multicolumn{3}{l}{Abell 1835}  &             &             &            &            &      \\
 305  & (1) &   1.13       &     1.21    &    1.24     &    1.26    &    1.28    &  1.30\\  
 319  &     &   1.15       &     1.26    &    1.29     &    1.31    &    1.33    &  1.36\\  
 347  &     &   1.17       &     1.28    &    1.32     &    1.34    &    1.36    &  1.39\\  
 532  &     &   1.33       &     1.60    &    1.71     &    1.77    &    1.83    &  1.91\\  
 676  &     &   1.34       &     1.62    &    1.75     &    1.81    &    1.89    &  1.98\\  
1093  & (2) &   1.43       &     1.82    &    1.99     &    2.09    &    2.20    &  2.34\\ 
      &     &              &             &             &            &      &      \\
 311  &(17) &   1.17       &     1.28    &    1.32     &    1.34    &    1.36    &  1.39\\  
 314  &(11) &   1.15       &     1.26    &    1.29     &    1.31    &    1.33    &  1.36\\  
 454  &(10) &   1.22       &     1.38    &    1.44     &    1.47    &    1.50    &  1.54\\  
 493  & (3) &   1.27       &     1.48    &    1.57     &    1.61    &    1.66    &  1.72\\ 
 504  & (4) &   1.29       &     1.51    &    1.60     &    1.64    &    1.70    &  1.76\\  
      &     &              &             &             &            &      &      \\
\multicolumn{3}{l}{AC114}  &             &             &            &            &      \\
 512  &     &   1.60       &     2.74    &    3.47     &    3.97    &    4.61    &  5.63\\  
 572  &     &   1.32       &     1.71    &    1.88     &    1.97    &    2.06    &  2.19\\  
 632  &     &   1.70       &     3.03    &    3.84     &    4.39    &    5.06    &  6.07\\  
 680  &     &   1.65       &     3.03    &    4.01     &    4.74    &    5.75    &  7.56\\  
 707  &     &   1.60       &     2.71    &    3.40     &    3.86    &    4.45    &  5.36\\  
 862  &     &   1.31       &     1.69    &    1.85     &    1.94    &    2.04    &  2.16\\  
1006  &     &   1.50       &     2.26    &    2.65     &    2.88    &    3.14    &  3.50\\  
1087  &     &   1.20       &     1.42    &    1.51     &    1.56    &    1.61    &  1.66\\ 
1167  & (1) &   1.24       &     1.53    &    1.64     &    1.71    &    1.77    &  1.86
    
\end{tabular}
\end{table*}


\subsection{Surface density}
\label{density}
\begin{figure}[h!]
   \centering \includegraphics[width=8cm]{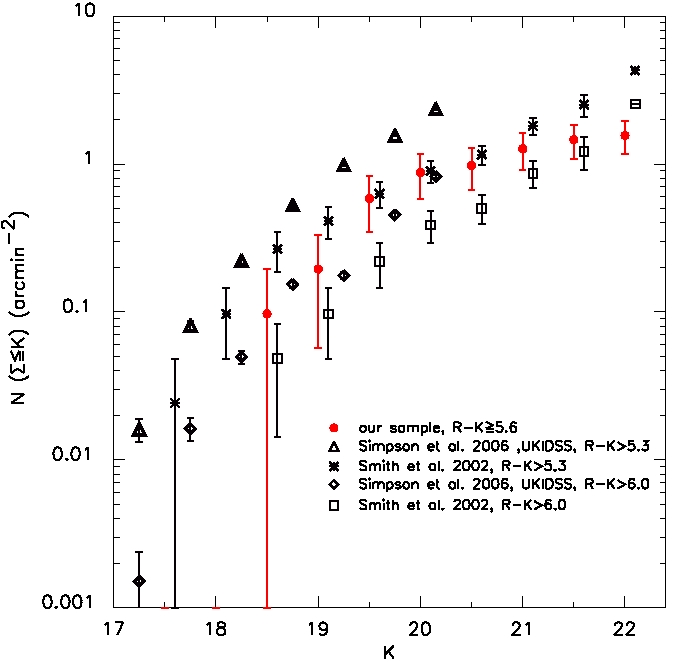}
   \caption{Cumulative surface density of our
     sample ($R-Ks\ge$ 5.6) after correcting for amplification (assuming a source
     plane at $z_{s}$=1.5) and incompleteness. We also include the ERO 
     sample ($R-K\ge$ 5.3 and 6.0) of \citet{smith2002} found in the
     fields of 10 massive galaxy cluster lenses and EROs found by \citet{simpson2006} in the UKIDSS Ultra Deep
     Survey Early Data Release ($R-K\ge$ 5.3).}
   \label{sigma}
   \end{figure}

After correction for lensing and incompleteness we obtain the surface density of EROs
shown in Fig.\ref{sigma}.
Although we use a slightly redder colour threshold than most authors, 
we compare our cumulative number counts with various other surveys. 
A comparable ERO sample in terms of environment is available from
\citet{smith2002}, who studied 10 massive galaxy
clusters at z$\sim$0.2, including Abell 1835. However, the single ERO
detected there has a colour of $R-K$=5.3 and hence does not satisfy our
colour threshold, while even the brightest of our EROs in Abell 1835 
(\#532 K=17.97) is not part
of this sample. The explanation lies both in the smaller field size of 
the UKIRT camera (1.5 arcmin) and the centering of the field on the
central cluster galaxy. As a result EROJ140057+0252.4 
\citep[see][]{smith2002} lies at the very edge of our image, which we
excluded due to the low signal-to-noise. The bright incompleteness
limits (80\% at 20.6$^{mag}$) in Smith et al. (\citeyear{smith2002})
might be responsible for their non-detection of our fainter
objects. 
The largest ERO survey, using the UKIDSS Ultra Deep survey, has been 
released recently by \citet{simpson2006}.
However, for now a comparison is only possible at brighter 
magnitudes, K$<$20.1.

Fig. \ref{sigma} shows the cumulative surface density of
$R-Ks\geq$5.6 EROs in comparison with the samples of \citet{smith2002}
and
\citet{simpson2006} ($R-K>5.3$ and $R-K>6$).\\
Up to $K$=20.5$^{mag}$, our number counts are slightly lower than those 
found by \citep{smith2002,smail2002} but agree well within the
1$\sigma$ error bars. Our lower number counts are also to be expected 
due to the redder colour threshold. We estimate the
cumulative surface density of EROs at $Ks\le$20.5 with (0.97$\pm$0.31) 
arcmin$^{-2}$, compared to  (1.16$\pm$0.17) arcmin$^{-2}$  and (0.50$\pm$0.11) arcmin$^{-2}$by
\citet{smith2002} for $R-K\ge$5.3 and $R-K\ge$6.0 (up to the same K-limit). The number counts increase
only slightly for fainter magnitudes, up to (1.36$\pm$0.36)
arcmin$^{-2}$ at $Ks\le$22.0.\\
The slope of N($\ge$ K)= 10$^{\alpha K}$ for our sample is $\alpha$=0.78$\pm$0.03 for
K$<$20.0, compared to $\alpha$=1.05$\pm$0.05 from \citet{daddi2000}
and $\alpha$=1.04$\pm$0.05 for \citet{smith2002}. This slope decreases for fainter magnitudes (20 $< K < $ 22) to
$\alpha$=0.11$\pm$0.01. Such a break in the cumulative number counts at
K$\sim$19-20 has been observed in
various surveys \citep{smith2002,daddi2000}, although the actual
values for $\alpha$ are found to be larger by a factor of $\sim$2-3 in 
\citep{smith2002,daddi2000} at fainter magnitudes ($K>$20.5). This
flattening  could be caused by the absence of evolved ellipticals with 
fainter magnitudes.\\
The preliminary results from UKIDSS UDS EDR \citep{simpson2006}
(K$\le$20.15) show much higher number counts than any other survey
\citep[e.g.][]{daddi2000,smail2002,smith2002}. The authors attribute this
result partly to the use of different filters and apertures. 

\section{Analysis of the observed SED}
\label{s_fit}

\subsection{SED fitting method}
To analyse quantitatively the observed SEDs we follow the procedures
outlined in \citet{schaerer2005} and described in detail in \citet{s07}.
We shall only briefly summarise the main points here.

The photometry in all bands except 24 \micron\ with MIPS has been used.
To take uncertainties in absolute flux calibrations between different instruments
into account we adopt a minimum photometric error of 0.1 mag in most computations.
We have used an updated version of the \hyperz\ code 
from \citet{bolzonella2000} to constrain the redshift, stellar population 
properties (age, star formation history), and extinction of the galaxies 
studied in this paper. To do so we use in particular a large library
of  synthetic, empirical and semi-empirical spectral templates.
The templates are gathered into the 4 following groups:\\
{\em 1)}
Bruzual \& Charlot plus \citet{coleman1980} empirical templates galaxies 
of all Hubble types (hereafter named BC or BCCWW group).
The theoretical Bruzual models \citep{bruzual2003}, taken here for solar 
metallicity, include various 
star formation histories representative of different Hubble types.
The IMF adopted in these models is the Miller-Scalo IMF from 0.1 to 125 \msun.\\
{\em 2)}
Starburst SEDs from Schaerer (\citeyear{schaerer2002,schaerer2003}) models at 
different metallicities extended up to ages of 1 Gyr and considering instantaneous 
bursts or constant star formation (hereafter s04gyr group).
These models assume a Salpeter IMF from 1 to 100 \msun.\\
{\em 3)}
Empirical or semi-empirical starburst, ULIRG and QSO templates.
We use starburst templates from the \citet{calzetti1994} and 
\citet{kinney1996} atlas and the HST QSO template of \citet{zheng1997}. 
To include also more obscured objects we have added 
UV to millimeter band templates of EROs, ULIRGS, starburst and normal 
galaxies (HR 10, Arp 220, M82, NGC 6090, M51, M100, NGC 6949) from fits of GRASIL models
to multi-wavelength observations (\citet{silva1998}, named GRASIL group). 
This template group will be used in particular to predict mid-IR to sub-mm fluxes,
and hence to estimate total bolometric luminosities,
after fitting the optical to 8 \micron\ part of the spectrum.
The main free parameters we consider are: the spectral template (among
a group), redshift $z$ , and (additional) extinction ($A_V$) assuming a
\citet{calzetti2000} law. 
To increase the diversity of empirical or semi-empirical templates
and to allow for possible deviations from them, we also allow
  for an additional reddening.
From the luminosity distance of the object or, if templates generated by 
evolutionary synthesis models are used, by scaling the template
SED to the observed absolute fluxes we obtained the absolute scaling
for properties such as stellar mass or the star formation rate (SFR).
In some cases we also use the bolometric luminosity computed from 
a GRASIL template to determine the SFR.
Finally, the absolute quantities  must also be corrected
for the effects of gravitational lensing. The magnification factors
listed in Table \ref{mag} are used for this purpose.\\

\subsection{Results}
\subsubsection{Abell 1835}
SED fits and the derived properties for the optical drop-out objects 
\#305 (1), \#311 (17), \#314 (11), \#454 (10), \#493 (3), \#504 (4), and \#1093 (2), 
corresponding to the objects of \citet{r06} with IDs in brackets, 
have already been discussed in depth in \citet{s07}. In case of
  \#1093 (2), the known sub-mm source SMMJ14009+0252, this includes
  also the SCUBA measurements by \citet{ivison00}.
We therefore limit the discussion here to the new objects in this field,
i.e.\  \#319, \#347, \#532, and \#676. 
A summary of their derived properties is given in Table \ref{t_props}.
For completeness and comparison the derived properties of the
objects from \citet{s07} are listed at the bottom of this table.
Note the stellar masses have been corrected by $(1+z)^{-1}$ to eliminate
an error in the absolute scaling found in \citet{s07}.
In contrast to the degeneracies found for many of the objects discussed
in \citet{s07} the photometric redshifts of the ``new'' objects show all 
well defined best fits at low-$z$ redshifts, $z \sim$ 0.9 to 2.5.
Object \#319, one of the two newly identified $I$ drop-outs, has the
highest photometric redshift, $\zfit \sim$ 2.4--2.5, which is well 
constrained by the ``curvature'' measured in the IRAC bands
due to the stellar peak at 1.6 \micron\ (restframe).
Except possibly for \#319, none of these objects are detected with MIPS 
at 24 \micron.

Object  \#319, is best fitted without extinction and with templates
of elliptical galaxies or maximally old simple stellar populations
(bursts). From SED fitting, this object is thus best characterised as ``elliptical''
in agreement with its empirical classification (Fig.\ref{eros1}). However, these SEDs are not able to explain the 24 \micron\
  flux.

The three remaining objects all show clear indications for dust, younger
ages, and short star formation histories (``bursts''), although they would
be classified as ``ellipticals'' according to their (R-K) vs (J-K) colours
(see Fig.\ \ref{eros1}).
The estimated stellar masses of all these objects are between 
$M_\star \times \mu \sim 3. \times 10^{10}$ and  $10^{12}$ \msun\ with
small magnification factors $\mu$. 

\begin{figure}[htb]
\centering \includegraphics[width=8.8cm]{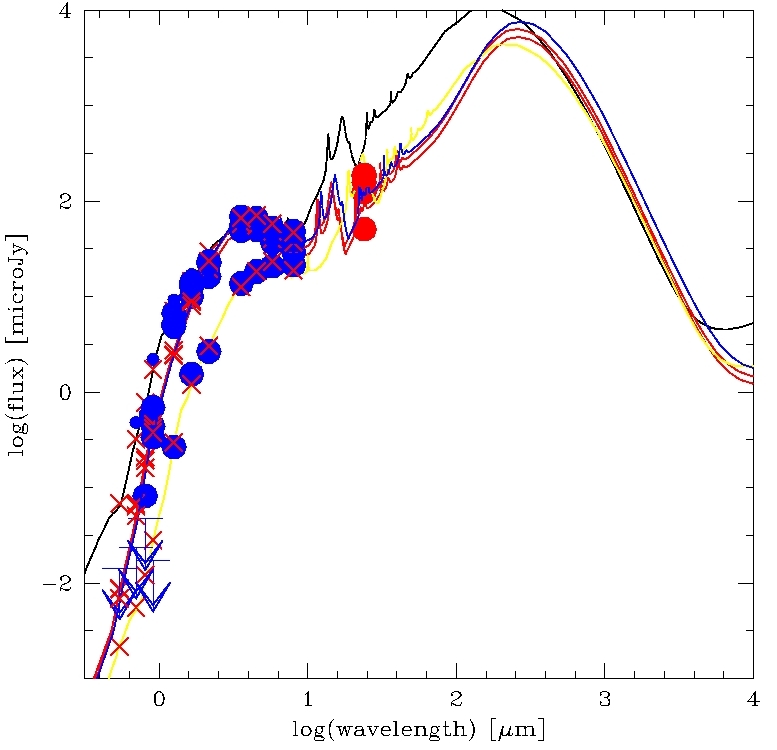}
\caption{SED fits with GRASIL templates for selected AC114 objects. Black: \#572,
red: \#1006, \#1087, yellow: \#1067 (highest-z), blue: \#1167. Sub-mm data (APEX/LABOCA) for this southern cluster are being aquired.}
\label{fig_eroac114}
\end{figure}

\subsubsection{AC114}
Using the templates from the BCCWW and s04gyr groups all except one object
have best-fit redshifts of $\zfit \sim$ 1.--2.6 (see Table \ref{t_props}).
The exception is the $R$-dropout \#1087, whose SED shows a very 
rapid and strong decline between the near-IR and optical bands, which is 
better fit with a Lyman-break than with a Balmer break. Formally its
best photometric redshift is thus $\zfit \sim 7.$, but given the
brightness of this object at near-IR and Spitzer bands the high-$z$
solution is extremely unlikely.
Using the semi-empirical GRASIL templates the best fit is found at $\zfit \sim 0.9$.
This solution also naturally explains the observed 24 \micron\ flux
of this object, as shown in Fig.\ \ref{fig_eroac114}.
For \#862 the best fit with the BCCWW and s04gyr templates is found at $\zfit
\sim$ 1.1, where GRASIL templates yield a somewhat lower redshift of $\sim$ 0.6
retained in the Table.

For 4 objects (\#512, \#632, \#680, \#707) the SED fitting indicates relatively
small amounts of extinction (\av $\la$ 0.6).
Interestingly these correspond precisely to all the objects which are
non-detected at 24 \micron. From their star formation history (all burst-like)
and their low (but not zero?) extinction these objects resemble ellipticals,
in agreement with their empirical classification.
For the brightest of all our EROs, \#632, we obtain different solutions
with the BC and S04gyr templates respectively, namely 
$\zfit \sim$ 1.3 (2.5), ages of 4.5 (2.6) Gyr, and little or no extinction, 
\av $\sim$ 0.6 (0.).
The estimated stellar masses of these objects are between 
$M_\star \times \mu \sim 6. \times 10^{10}$ and  $10^{12}$ \msun.

The remaining objects, also all detected at 24 \micron, show all clear 
evidence for substantial extinction. In Table \ref{t_props} we list
their properties derived from fits with the GRASIL templates.
The best fits are found with the M51 or M82 templates requiring, however,
additional extinction in 5 of 6 cases.
The predicted 24 \micron\ fluxes,
not included in the SED modelling, agree well with the observations
except maybe for 572 whose MIPS flux is somewhat overestimated
(see Fig. \ref{fig_eroac114}).
In conclusion, for 572, 862, 1006, 1087, and 1167 all indications
(SED fitting and MIPS detections) consistently confirm the dusty low-$z$
($\sim$ 0.9 to 2 or maybe 2.5) starburst nature of these objects. 
This shows the limitations of empirical diagrams, which would e.g.\
clearly classify three of them as ellipticals (cf.\ Fig.\ref{eros1}).
As already mentioned above, our conclusion does also not support 
the explanation of objects with such extreme $I-K$ colours 
(cf.\ Fig.\ref{IK45}) as ``old and dead galaxies'' as proposed 
by \citet{labbe05}.

\begin{table*}[htb]
\caption{Derived/estimated properties for EROs in Abell 1835 and AC114.
Listed are the object ID (col.\ 1), the photometric redshift estimate (col.\ 2), the extinction (col.\ 3),
the type of the best-fit template (col.\ 4), the distance modulus corresponding to \zphot\ (col.\ 5),
the absolute $Ks$-band magnitude non-corrected for lensing (col.\ 6), 
the absolute rest-frame $Ks$-band magnitude non-corrected for lensing (col.\ 7),
the estimated stellar mass (from scaling the SED fit or from $M^{\rm rest}(Ks)$ assuming
$L_K/M=3.2$, col.\ 8), the estimated star formation rate  non-corrected for lensing (col.\ 9),
and the age of the stellar population (col.\ 10).
To correct the above mentioned absolute quantities for gravitational magnification the appropriate
magnification factors listed in Table \ref{mag} must be used.
The data at the bottom of the Table is from \citet{s07} (Tabs.\ 5 and 6), except
for the stellar mass, which has been corrected by $(1+z)^{-1}$ to eliminate
an error in that paper.}

\begin{tabular}{rrlllllllllll}
\hline
\hline
Object & &\zphot\ & \av\ & template & DM$^a$ & $M^{\rm rest}(Ks)-2.5\log(\mu)$ & Mass$\times \mu$ & SFR$\times \mu$ & stellar age  \\
       & &        & [mag]&          & [mag]  & [mag]  & \msun& \msunyr & [Gyr] \\
\hline
\multicolumn{5}{l}{Abell 1835:} \\
298 & &0.8--1.1 & 0        & elliptical & 43.65 & -22.0  \\
319 & &2.4--2.5 & $\sim 0$ & burst      & 46.60 & -26.4 & 3.3$\times 10^{11}$ &  & 2.3 \\
347 & &1.1--1.2 & 1.6--2.2 & burst      & 44.35 & -24.4 & (3.1-4.4)$\times 10^{10}$ &  & 0.5--1.0 \\
532 & &1.3--1.4 & 0.4--2.0 & burst      & 44.71 & -26.4 & (3.0-4.6)$\times 10^{11}$ &  & 1.0--3.5 \\
676 & &$\sim$ 1.4 & 1.4--1.8 & burst    & 45.0  & -24.8 & (3.3-3.8)$\times 10^{10}$ &  & 0.4--0.7 \\
\multicolumn{5}{l}{AC114:} \\
512 & &2.4--2.6 & 0.4      &  burst     & 46.65 & -25.2 & 4.3$\times 10^{10}$ &  & 0.7 \\
572 & &$\sim$ 1.2  &       & M82   & 44.65 & -26.1 &     & 35.7 &    \\
632 & &$\sim$ 1.3 (2.5)  & 0.6 (0.)  & burst      & 44.76 & -27.5 & 1.5$\times 10^{12}$ &  & 4.5 (2.6) \\
680 & &$\sim$ 2.1  & 0.6   & burst      & 46.12 & -26.5 & 2.3$\times 10^{11}$ &  & 1.4 \\
707 & &$\sim$ 1.2  & 0.6   & elliptical & 44.63 & -25.8 &     &  &     \\
862 & &$\sim$ 0.6  & +3.2  & M51   & 42.76 & -24.2 &     & 10.6  & \\    
1006& &$\sim$ 0.9  & +3.6  & M51   & 43.86 & -25.4 &     & 30.5  & \\
1067& &$\sim$ 2.0  & +2.0  & M82   & 46.02 & -26.0 &     & 205.5 & \\
1087& &$\sim$ 0.9  & +3.8  & M51   & 43.77 & -25.6 &     & 34.2  & \\
\hline
305 &(1)  & $\sim$ 0.4 --1.5    & ?       &    \multicolumn{5}{l}{Fits uncertain -- see paper I} \\
311 &(17) & $\sim$ 0.7--0.8 & $\sim$ 3.8  & burst     & 43.0 & -21.7 & $\sim 7.6 \times 10^{9}$? & $\sim$ 0.9 &  ? (see paper I)\\  
504 &(4)  & $\sim$ 1.2    & 0--1.6  & burst/elliptical & 44.60 & -21.6 & $\sim 7.7 \times 10^{9}$ & $\sim$ 5 & 0.7 to 4.5 \\
1093 &(2)  & $\sim$ 2.8--3 & 2.4--3  & young burst & 47.0 & -27.7 & $\sim 3.2 \times 10^{11}$ & $\sim$2100 & $<$ 0.36 \\
1167 &(1)  & $\sim$ 1.3--1.6& $\sim$ 1.6--2.8 & burst & 44.84 & -26.4 & $(0.6-1.1) \times 10^{12}$ &  & $\sim$ 0.9--4.5 Gyr \\
1167 &(1)  & $\sim$ 1.0     & +3.8      & M51         & 44.03 & -25.9 &                      & $\sim$ 48 & \\
\\
 493 &(3)  & $\sim$ 1.1 & $\sim$ 0.6--0.8 & burst & 44.4 & -22.2 & $\sim 2.4 \times 10^{9}$ & & 0.5 \\
 454  &(10) & $\sim$ 1.2 & $\sim$ 1.8 & burst & 44.68 & -22.9 & $\sim 3.6 \times 10^{9}$ & & 0.5 \\ 
 314  &(11) &  ?  \\
\hline
 \multicolumn{5}{l}{$^a$ distance modulus computed for minimum redshift}
\end{tabular}
\label{t_props}
\end{table*}

The star formation rates estimated for these dusty objects from the 
bolometric luminosity of the GRASIL model fit are between $SFR \sim$ 
15 and 120 \msunyr, after correction for lensing.
Their bolometric luminosities classify them in the range of luminous
infrared galaxies (LIRG) with $L_{\rm bol} > 10^{11}$ \lsun.
\subsection{Discussion}
As we can see from Table \ref{t_props} the properties of our EROs 
span a rather wide range in extinction, stellar age, and stellar mass.
The properties and their range are quite similar to those determined
for DRGs by \citet{foerster2004} and for
the IEROs of \citet{yan04} in \citet{s07}.
However, in our sample we find some extreme objects in terms of colours,
for which the SED modelling indicates quite clearly very high extinction
($A_V \sim 3$ and higher), which is not found in the DRG samples 
of \citet{foerster2004} and \citet{papovich06}.
According to our analysis \citep[see][]{s07} some IEROs of \citet{yan04}
show also such high extinction.

\section{Summary and conclusions}
\label{s_conclude}
We have combined new ACS/HST observations, Spitzer IRAC and MIPS guaranteed
time observations, and the optical and near-IR observations
of \citet{r06} of two well-known lensing clusters, Abell 1835 and AC114, 
to study extremely red galaxies (EROs) in these fields.
New and archival X-ray observations with ACIS/Chandra have also been 
obtained for these clusters.

Using a standard $R-K\ge 5.6$ criterion we have found 6 and 9
EROs in Abell 1835 and AC114 respectively.
Several (8) of these objects are undetected up to the $I$ and/or $z_{850}$
band, and are hence ``optical'' drop-out sources. Three of them,
already identified earlier by \citet{r06}, have been discussed in detail
in \citet{s07}.

We have discussed the empirical properties of these EROs and compared them
to other samples in the literature. We have also undertaken SED modelling
based on a modified version of the \hyperz\ photometric redshift code 
and using a large number of spectral templates, including also very dusty galaxies.

The main results can be summarised as follows:
\begin{itemize}
\item Among our EROs we find 3 sources showing quite unusually 
red colours in $R-K$ and other colours. Few similar objects are found e.g.
among the samples of IRAC selected IEROs of \citet{yan04}, EROs of \citet{sawicki2005} and the 
DRGs of \citet{labbe05}.
Our source density is compatible with other counts from the literature.

\item After correcting for lens amplification, we estimate a surface
  density of (0.97$\pm$0.31) arcmin$^{-2}$ for EROs with ($R-K \ge
  $5.6) at K$<20.5$. We observe a significant flattening of the
  number count  at K$\sim$20, possibly the result of loosing the
  contribution of bright evolved ellipticals to the overall ERO
  population.

\item According to ``empirical'' and to \hyperz\ modelling, the photometric redshifts 
of most of our sources yield are $z\sim$ 0.7--1.5. Five of them are found
at higher redshift ($z \sim$ 2.--2.5.).

\item According to simple colour-colour diagrams the majority of
our objects would be classified as hosting old stellar populations (``ellipticals'').
However, there are clear signs of dusty starbursts for several among them.
These objects correspond to the most extreme ones in $R-K$ colour.

\item We found that some very red DRGs, which would be classified as old and dead galaxies 
according to other studies (e.g.\ \citet{labbe05}), are rather 
very dusty starbursts, even (U)LIRGs, as also supported their mid-IR photometry.
Estimates of the fraction of old and dead galaxies among red galaxies
may thus need to be treated with caution.

\item As in earlier studies an overlap of different populations is found.
Among our 15 EROs six also classify as DRGs (40\%).
12 of 14 EROs (85 \%) with available IRAC photometry also fulfil
the selection criteria for IRAC selected IEROs of \citet{yan04}.
Objects which do not classify as IERO are also not DRGs; the reverse
is however not true.
SED modelling shows that $\sim$ 40 \% of the IEROs are luminous
or ultra-luminous infrared galaxies ((U)LIRG).

\item None of our objects detected at X-rays above the cluster background
emission, with upper limits typically of the order of $\sim (3-10) \times
10^{-16}$ ergs s$^{-1}$cm$^{-2}$ in the 0.5--7.0 keV band.
No indication for AGNs is found, although faint activity cannot be excluded
for all objects. From mid-IR and X-ray data 5 objects are clearly classified 
as starbursts.

\item Quantitative SED fitting for our objects shows that
they cover a fairly wide range in properties, such as extinction,
stellar age, mass, and SFR. The derived properties are quite similar 
to those of DRGs and IEROs, except for 5 extreme objects in terms of colours,
for which a very high extinction ($A_V \ga 3$) is found.
According to our analysis some IEROs of \citet{yan04}
show also such high extinction \citep[see][]{s07}.
From the SED modelling these 5 EROs are expected to be 
(U)LIRG, and their IR to sub-mm SED is predicted.
\end{itemize}

Understanding the links between these different galaxy populations
and their evolutionary history remains largely to be done.\\

\begin{acknowledgements}

We thank  Andrea Grazian for providing us with the GOODS-MUSIC
catalogues and Graham Smith for making their results available
in electronic format.\\
Support from {\em ISSI} (International Space Science Institute) in Bern for an 
``International Team'' is kindly acknowledged. 
Part of this work was supported by the
Swiss National Science Foundation,
the French {\it Centre National de la Recherche Scientifique},
and the French {\it Programme National de
Cosmologie} (PNC) and {\it Programme National de Galaxies} (PNG).\\
This paper is based on observations collected at the European Space Observatory, Chile
(069.A-0508,070.A-0355,073.A-0471), and the Canada-France-Hawaii Telescope operated by
the National Research Council of Canada, the French Centre National
de la Recherche Scientifique (CNRS) and the University of Hawaii,
and the NASA/ESA Hubble Space Telescope operated by the Association of 
Universities for Research in Astronomy, Inc., 
the Spitzer Space Telescope, which is operated by the Jet Propulsion Laboratory, 
California Institute of Technology under NASA contract 1407,
and the Chandra satellite.
This research was supported in part by Chandra General Observer
Program grant GO6-7106X.

\end{acknowledgements}

\bibliographystyle{aa} 
\bibliography{hempel}   

\begin{thebibliography}{86}
\expandafter\ifx\csname natexlab\endcsname\relax\def\natexlab#1{#1}\fi

\bibitem[{{Alexander} {et~al.}(2003){Alexander}, {Bauer}, {Brandt},
  {Schneider}, {Hornschemeier}, {Vignali}, {Barger}, {Broos}, {Cowie},
  {Garmire}, {Townsley}, {Bautz}, {Chartas}, \& {Sargent}}]{alexander2003}
{Alexander}, D.~M., {Bauer}, F.~E., {Brandt}, W.~N., {et~al.} 2003, \aj, 126,
  539

\bibitem[{{Alexander} {et~al.}(2002){Alexander}, {Vignali}, {Bauer}, {Brandt},
  {Hornschemeier}, {Garmire}, \& {Schneider}}]{alexander02}
{Alexander}, D.~M., {Vignali}, C., {Bauer}, F.~E., {et~al.} 2002, \aj, 123,
  1149

\bibitem[{{Alonso-Herrero} {et~al.}(2006){Alonso-Herrero},
  {P{\'e}rez-Gonz{\'a}lez}, {Alexander}, {Rieke}, {Rigopoulou}, {Le Floc'h},
  {Barmby}, {Papovich}, {Rigby}, {Bauer}, {Brandt}, {Egami}, {Willner}, {Dole},
  \& {Huang}}]{alonsoherr2006}
{Alonso-Herrero}, A., {P{\'e}rez-Gonz{\'a}lez}, P.~G., {Alexander}, D.~M.,
  {et~al.} 2006, \apj, 640, 167

\bibitem[{{Alonso-Herrero} {et~al.}(2004){Alonso-Herrero},
  {P{\'e}rez-Gonz{\'a}lez}, {Rigby}, {Rieke}, {Le Floc'h}, {Barmby}, {Page},
  {Papovich}, {Dole}, {Egami}, {Huang}, {Rigopoulou},
  {Crist{\'o}bal-Hornillos}, {Eliche-Moral}, {Balcells}, {Prieto}, {Erwin},
  {Engelbracht}, {Gordon}, {Werner}, {Willner}, {Fazio}, {Frayer}, {Hines},
  {Kelly}, {Latter}, {Misselt}, {Miyazaki}, {Morrison}, {Rieke}, \&
  {Wilson}}]{alonsoherr2004}
{Alonso-Herrero}, A., {P{\'e}rez-Gonz{\'a}lez}, P.~G., {Rigby}, J., {et~al.}
  2004, \apjs, 154, 155

\bibitem[{{Bergstr{\"o}m} \& {Wiklind}(2004)}]{wiklind04}
{Bergstr{\"o}m}, S. \& {Wiklind}, T. 2004, \aap, 414, 95

\bibitem[{{Bertin} \& {Arnouts}(1996)}]{bertin96}
{Bertin}, E. \& {Arnouts}, S. 1996, \aaps, 117, 393

\bibitem[{{Bolzonella} {et~al.}(2000){Bolzonella}, {Miralles}, \&
  {Pell{\'o}}}]{bolzonella2000}
{Bolzonella}, M., {Miralles}, J.-M., \& {Pell{\'o}}, R. 2000, \aap, 363, 476

\bibitem[{{Brusa} {et~al.}(2005){Brusa}, {Comastri}, {Daddi}, {Pozzetti},
  {Zamorani}, {Vignali}, {Cimatti}, {Fiore}, {Mignoli}, {Ciliegi}, \&
  {R{\"o}ttgering}}]{brusa2005}
{Brusa}, M., {Comastri}, A., {Daddi}, E., {et~al.} 2005, \aap, 432, 69

\bibitem[{{Bruzual} \& {Charlot}(2003)}]{bruzual2003}
{Bruzual}, G. \& {Charlot}, S. 2003, \mnras, 344, 1000

\bibitem[{{Calzetti} {et~al.}(2000){Calzetti}, {Armus}, {Bohlin}, {Kinney},
  {Koornneef}, \& {Storchi-Bergmann}}]{calzetti2000}
{Calzetti}, D., {Armus}, L., {Bohlin}, R.~C., {et~al.} 2000, \apj, 533, 682

\bibitem[{{Calzetti} {et~al.}(1994){Calzetti}, {Kinney}, \&
  {Storchi-Bergmann}}]{calzetti1994}
{Calzetti}, D., {Kinney}, A.~L., \& {Storchi-Bergmann}, T. 1994, \apj, 429, 582

\bibitem[{{Campusano} {et~al.}(2001){Campusano}, {Pell{\'o}}, {Kneib}, {Le
  Borgne}, {Fort}, {Ellis}, {Mellier}, \& {Smail}}]{campusano2001}
{Campusano}, L.~E., {Pell{\'o}}, R., {Kneib}, J.-P., {et~al.} 2001, \aap, 378,
  394

\bibitem[{{Cimatti} {et~al.}(1999){Cimatti}, {Daddi}, {di Serego Alighieri},
  {Pozzetti}, {Mannucci}, {Renzini}, {Oliva}, {Zamorani}, {Andreani}, \&
  {R{\"o}ttgering}}]{cimatti99}
{Cimatti}, A., {Daddi}, E., {di Serego Alighieri}, S., {et~al.} 1999, \aap,
  352, L45

\bibitem[{{Cimatti} {et~al.}(2002){Cimatti}, {Pozzetti}, {Mignoli}, {Daddi},
  {Menci}, {Poli}, {Fontana}, {Renzini}, {Zamorani}, {Broadhurst}, {Cristiani},
  {D'Odorico}, {Giallongo}, \& {Gilmozzi}}]{cimatti02b}
{Cimatti}, A., {Pozzetti}, L., {Mignoli}, M., {et~al.} 2002, \aap, 391, L1

\bibitem[{{Coleman} {et~al.}(1980){Coleman}, {Wu}, \& {Weedman}}]{coleman1980}
{Coleman}, G.~D., {Wu}, C.-C., \& {Weedman}, D.~W. 1980, \apjs, 43, 393

\bibitem[{{Cowie} {et~al.}(1994){Cowie}, {Gardner}, {Hu}, {Songaila}, {Hodapp},
  \& {Wainscoat}}]{cowie94}
{Cowie}, L.~L., {Gardner}, J.~P., {Hu}, E.~M., {et~al.} 1994, \apj, 434, 114

\bibitem[{{Czoske} {et~al.}(2003){Czoske}, {Kneib}, \& {Bardeau}}]{czoske2003}
{Czoske}, O., {Kneib}, J.-P., \& {Bardeau}, S. 2003, in Astronomical Society of
  the Pacific Conference Series, Vol. 301, Astronomical Society of the Pacific
  Conference Series, ed. S.~{Bowyer} \& C.-Y. {Hwang}, 281--+

\bibitem[{{Daddi} {et~al.}(2002){Daddi}, {Cimatti}, {Broadhurst}, {Renzini},
  {Zamorani}, {Mignoli}, {Saracco}, {Fontana}, {Pozzetti}, {Poli}, {Cristiani},
  {D'Odorico}, {Giallongo}, {Gilmozzi}, \& {Menci}}]{daddi02}
{Daddi}, E., {Cimatti}, A., {Broadhurst}, T., {et~al.} 2002, \aap, 384, L1

\bibitem[{{Daddi} {et~al.}(2000){Daddi}, {Cimatti}, {Pozzetti}, {Hoekstra},
  {R{\"o}ttgering}, {Renzini}, {Zamorani}, \& {Mannucci}}]{daddi2000}
{Daddi}, E., {Cimatti}, A., {Pozzetti}, L., {et~al.} 2000, \aap, 361, 535

\bibitem[{{De Filippis} {et~al.}(2004){De Filippis}, {Bautz}, {Sereno}, \&
  {Garmire}}]{defilippis2004}
{De Filippis}, E., {Bautz}, M.~W., {Sereno}, M., \& {Garmire}, G.~P. 2004,
  \apj, 611, 164

\bibitem[{{Della Ceca} {et~al.}(2004){Della Ceca}, {Maccacaro}, {Caccianiga},
  {Severgnini}, {Braito}, {Barcons}, {Carrera}, {Watson}, {Tedds}, {Brunner},
  {Lehmann}, {Page}, {Lamer}, \& {Schwope}}]{della2004}
{Della Ceca}, R., {Maccacaro}, T., {Caccianiga}, A., {et~al.} 2004, \aap, 428,
  383

\bibitem[{{Egami} {et~al.}(2006){Egami}, {Misselt}, {Rieke}, {Wise},
  {Neugebauer}, {Kneib}, {Le Floc'h}, {Smith}, {Blaylock}, {Dole}, {Frayer},
  {Huang}, {Krause}, {Papovich}, {P{\'e}rez-Gonz{\'a}lez}, \&
  {Rigby}}]{Egami2006}
{Egami}, E., {Misselt}, K.~A., {Rieke}, G.~H., {et~al.} 2006, \apj, 647, 922

\bibitem[{{Elston} {et~al.}(1988){Elston}, {Rieke}, \& {Rieke}}]{elston88}
{Elston}, R., {Rieke}, G.~H., \& {Rieke}, M.~J. 1988, \apjl, 331, L77

\bibitem[{{Elston} {et~al.}(1989){Elston}, {Rieke}, \& {Rieke}}]{elston89}
{Elston}, R., {Rieke}, M.~J., \& {Rieke}, G.~H. 1989, \apj, 341, 80

\bibitem[{{Fazio} {et~al.}(2004){Fazio}, {Hora}, {Allen}, {Ashby}, {Barmby},
  {Deutsch}, {Huang}, {Kleiner}, {Marengo}, {Megeath}, {Melnick}, {Pahre},
  {Patten}, {Polizotti}, {Smith}, {Taylor}, {Wang}, {Willner}, {Hoffmann},
  {Pipher}, {Forrest}, {McMurty}, {McCreight}, {McKelvey}, {McMurray}, {Koch},
  {Moseley}, {Arendt}, {Mentzell}, {Marx}, {Losch}, {Mayman}, {Eichhorn},
  {Krebs}, {Jhabvala}, {Gezari}, {Fixsen}, {Flores}, {Shakoorzadeh}, {Jungo},
  {Hakun}, {Workman}, {Karpati}, {Kichak}, {Whitley}, {Mann}, {Tollestrup},
  {Eisenhardt}, {Stern}, {Gorjian}, {Bhattacharya}, {Carey}, {Nelson},
  {Glaccum}, {Lacy}, {Lowrance}, {Laine}, {Reach}, {Stauffer}, {Surace},
  {Wilson}, {Wright}, {Hoffman}, {Domingo}, \& {Cohen}}]{Fazio2004}
{Fazio}, G.~G., {Hora}, J.~L., {Allen}, L.~E., {et~al.} 2004, \apjs, 154, 10

\bibitem[{{Ferguson} {et~al.}(2000){Ferguson}, {Dickinson}, \&
  {Williams}}]{ferguson2000}
{Ferguson}, H.~C., {Dickinson}, M., \& {Williams}, R. 2000, \araa, 38, 667

\bibitem[{{Fiore} {et~al.}(2003){Fiore}, {Brusa}, {Cocchia}, {Baldi},
  {Carangelo}, {Ciliegi}, {Comastri}, {La Franca}, {Maiolino}, {Matt},
  {Molendi}, {Mignoli}, {Perola}, {Severgnini}, \& {Vignali}}]{fiore2003}
{Fiore}, F., {Brusa}, M., {Cocchia}, F., {et~al.} 2003, \aap, 409, 79

\bibitem[{{Fontana} {et~al.}(2004){Fontana}, {Pozzetti}, {Donnarumma},
  {Renzini}, {Cimatti}, {Zamorani}, {Menci}, {Daddi}, {Giallongo}, {Mignoli},
  {Perna}, {Salimbeni}, {Saracco}, {Broadhurst}, {Cristiani}, {D'Odorico}, \&
  {Gilmozzi}}]{fontana04}
{Fontana}, A., {Pozzetti}, L., {Donnarumma}, I., {et~al.} 2004, \aap, 424, 23

\bibitem[{{F{\"o}rster-Schreiber} {et~al.}(2004){F{\"o}rster-Schreiber}, {van
  Dokkum}, {Franx}, {Labb{\'e}}, {Rudnick}, {Daddi}, {Illingworth}, {Kriek},
  {Moorwood}, {Rix}, {R{\"o}ttgering}, {Trujillo}, {van der Werf}, {van
  Starkenburg}, \& {Wuyts}}]{foerster2004}
{F{\"o}rster-Schreiber}, N.~M., {van Dokkum}, P.~G., {Franx}, M., {et~al.}
  2004, \apj, 616, 40

\bibitem[{{Georgakakis} {et~al.}(2006){Georgakakis}, {Hopkins}, {Afonso},
  {Sullivan}, {Mobasher}, \& {Cram}}]{georga06}
{Georgakakis}, A., {Hopkins}, A.~M., {Afonso}, J., {et~al.} 2006, \mnras, 140

\bibitem[{{Gilbank} {et~al.}(2003){Gilbank}, {Smail}, {Ivison}, \&
  {Packham}}]{gilbank03}
{Gilbank}, D.~G., {Smail}, I., {Ivison}, R.~J., \& {Packham}, C. 2003, \mnras,
  346, 1125

\bibitem[{{Graham} \& {Dey}(1996)}]{graham96}
{Graham}, J.~R. \& {Dey}, A. 1996, \apj, 471, 720

\bibitem[{{Grazian} {et~al.}(2006){Grazian}, {Fontana}, {de Santis}, {Nonino},
  {Salimbeni}, {Giallongo}, {Cristiani}, {Gallozzi}, \&
  {Vanzella}}]{grazian2006}
{Grazian}, A., {Fontana}, A., {de Santis}, C., {et~al.} 2006, \aap, 449, 951

\bibitem[{{Grazian} {et~al.}(2007){Grazian}, {Nonino}, \&
  {Gallozzi}}]{grazian07}
{Grazian}, A., {Nonino}, M., \& {Gallozzi}, S. 2007, astro-ph/0701233

\bibitem[{{Houck} \& {Denicola}(2000)}]{houck2000}
{Houck}, J.~C. \& {Denicola}, L.~A. 2000, in ASP Conf. Ser. 216: Astronomical
  Data Analysis Software and Systems IX, ed. N.~{Manset}, C.~{Veillet}, \&
  D.~{Crabtree}, 591--+

\bibitem[{{Houck} {et~al.}(2005){Houck}, {Soifer}, {Weedman}, {Higdon},
  {Higdon}, {Herter}, {Brown}, {Dey}, {Jannuzi}, {Le Floc'h}, {Rieke}, {Armus},
  {Charmandaris}, {Brandl}, \& {Teplitz}}]{houck2005}
{Houck}, J.~R., {Soifer}, B.~T., {Weedman}, D., {et~al.} 2005, \apjl, 622, L105

\bibitem[{{Hu} \& {Ridgway}(1994)}]{hu94}
{Hu}, E.~M. \& {Ridgway}, S.~E. 1994, \aj, 107, 1303

\bibitem[{{Im} {et~al.}(2002){Im}, {Simard}, {Faber}, {Koo}, {Gebhardt},
  {Willmer}, {Phillips}, {Illingworth}, {Vogt}, \& {Sarajedini}}]{im02}
{Im}, M., {Simard}, L., {Faber}, S.~M., {et~al.} 2002, \apj, 571, 136

\bibitem[{{Ivison} {et~al.}(2000){Ivison}, {Smail}, {Barger}, {Kneib}, {Blain},
  {Owen}, {Kerr}, \& {Cowie}}]{ivison00}
{Ivison}, R.~J., {Smail}, I., {Barger}, A.~J., {et~al.} 2000, \mnras, 315, 209

\bibitem[{{Ivison} {et~al.}(2001){Ivison}, {Smail}, {Frayer}, {Kneib}, \&
  {Blain}}]{ivison2001}
{Ivison}, R.~J., {Smail}, I., {Frayer}, D.~T., {Kneib}, J.-P., \& {Blain},
  A.~W. 2001, \apjl, 561, L45

\bibitem[{{Kauffmann} {et~al.}(1993){Kauffmann}, {White}, \&
  {Guiderdoni}}]{kauffmann93}
{Kauffmann}, G., {White}, S.~D.~M., \& {Guiderdoni}, B. 1993, \mnras, 264, 201

\bibitem[{{Kinney} {et~al.}(1996){Kinney}, {Calzetti}, {Bohlin}, {McQuade},
  {Storchi-Bergmann}, \& {Schmitt}}]{kinney1996}
{Kinney}, A.~L., {Calzetti}, D., {Bohlin}, R.~C., {et~al.} 1996, \apj, 467, 38

\bibitem[{{Kitzbichler} \& {White}(2006)}]{kitzbichler06}
{Kitzbichler}, M.~G. \& {White}, S.~D.~M. 2006, \mnras, 366, 858

\bibitem[{{Kriek} {et~al.}(2006){Kriek}, {van Dokkum}, {Franx}, {Quadri},
  {Gawiser}, {Herrera}, {Illingworth}, {Labb{\'e}}, {Lira}, {Marchesini},
  {Rix}, {Rudnick}, {Taylor}, {Toft}, {Urry}, \& {Wuyts}}]{kriek2006}
{Kriek}, M., {van Dokkum}, P.~G., {Franx}, M., {et~al.} 2006, \apjl, 649, L71

\bibitem[{{Labb{\'e}} {et~al.}(2005){Labb{\'e}}, {Huang}, {Franx}, {Rudnick},
  {Barmby}, {Daddi}, {van Dokkum}, {Fazio}, {Schreiber}, {Moorwood}, {Rix},
  {R{\"o}ttgering}, {Trujillo}, \& {van der Werf}}]{labbe05}
{Labb{\'e}}, I., {Huang}, J., {Franx}, M., {et~al.} 2005, \apjl, 624, L81

\bibitem[{{Longhetti} {et~al.}(2005){Longhetti}, {Saracco}, {Severgnini},
  {Della Ceca}, {Braito}, {Mannucci}, {Bender}, {Drory}, {Feulner}, \&
  {Hopp}}]{longhetti05}
{Longhetti}, M., {Saracco}, P., {Severgnini}, P., {et~al.} 2005, \mnras, 361,
  897

\bibitem[{{Maccacaro} {et~al.}(1988){Maccacaro}, {Gioia}, {Wolter}, {Zamorani},
  \& {Stocke}}]{maccacaro1988}
{Maccacaro}, T., {Gioia}, I.~M., {Wolter}, A., {Zamorani}, G., \& {Stocke},
  J.~T. 1988, \apj, 326, 680

\bibitem[{{Magliocchetti} {et~al.}(2007){Magliocchetti}, {Silva}, {Lapi}, {de
  Zotti}, {Granato}, {Fadda}, \& {Danese}}]{magloocc2007}
{Magliocchetti}, M., {Silva}, L., {Lapi}, A., {et~al.} 2007, \mnras, 375, 1121

\bibitem[{{Mainieri} {et~al.}(2002){Mainieri}, {Bergeron}, {Hasinger},
  {Lehmann}, {Rosati}, {Schmidt}, {Szokoly}, \& {Della Ceca}}]{mainieri2002}
{Mainieri}, V., {Bergeron}, J., {Hasinger}, G., {et~al.} 2002, \aap, 393, 425

\bibitem[{{McCarthy}(2004)}]{mccarthy2004}
{McCarthy}, P.~J. 2004, \araa, 42, 477

\bibitem[{{Mignoli} {et~al.}(2004){Mignoli}, {Pozzetti}, {Comastri}, {Brusa},
  {Ciliegi}, {Cocchia}, {Fiore}, {La Franca}, {Maiolino}, {Matt}, {Molendi},
  {Perola}, {Puccetti}, {Severgnini}, \& {Vignali}}]{mignoli2004}
{Mignoli}, M., {Pozzetti}, L., {Comastri}, A., {et~al.} 2004, \aap, 418, 827

\bibitem[{{Moustakas} {et~al.}(2004){Moustakas}, {Casertano}, {Conselice},
  {Dickinson}, {Eisenhardt}, {Ferguson}, {Giavalisco}, {Grogin}, {Koekemoer},
  {Lucas}, {Mobasher}, {Papovich}, {Renzini}, {Somerville}, \&
  {Stern}}]{moustakas04}
{Moustakas}, L.~A., {Casertano}, S., {Conselice}, C.~J., {et~al.} 2004, \apjl,
  600, L131

\bibitem[{{Natarajan} {et~al.}(1998){Natarajan}, {Kneib}, {Smail}, \&
  {Ellis}}]{natarajan1998}
{Natarajan}, P., {Kneib}, J.-P., {Smail}, I., \& {Ellis}, R.~S. 1998, \apj,
  499, 600

\bibitem[{{Papovich}(2006)}]{papovich06}
{Papovich}, C. 2006, New Astronomy Review, 50, 134

\bibitem[{{Pettini} {et~al.}(2000){Pettini}, {Steidel}, {Adelberger},
  {Dickinson}, \& {Giavalisco}}]{pettini2000}
{Pettini}, M., {Steidel}, C.~C., {Adelberger}, K.~L., {Dickinson}, M., \&
  {Giavalisco}, M. 2000, \apj, 528, 96

\bibitem[{{Piccinotti} {et~al.}(1982){Piccinotti}, {Mushotzky}, {Boldt},
  {Holt}, {Marshall}, {Serlemitsos}, \& {Shafer}}]{piccinotti1982}
{Piccinotti}, G., {Mushotzky}, R.~F., {Boldt}, E.~A., {et~al.} 1982, \apj, 253,
  485

\bibitem[{{Pozzetti} {et~al.}(1996){Pozzetti}, {Bruzual A.}, \&
  {Zamorani}}]{pozzetti96}
{Pozzetti}, L., {Bruzual A.}, G., \& {Zamorani}, G. 1996, \mnras, 281, 953

\bibitem[{{Pozzetti} \& {Mannucci}(2000)}]{pozzetti00}
{Pozzetti}, L. \& {Mannucci}, F. 2000, \mnras, 317, L17

\bibitem[{{Ranalli} {et~al.}(2003){Ranalli}, {Comastri}, \&
  {Setti}}]{ranalli2003}
{Ranalli}, P., {Comastri}, A., \& {Setti}, G. 2003, \aap, 399, 39

\bibitem[{{Richard} {et~al.}(2006){Richard}, {Pell{\'o}}, {Schaerer}, {Le
  Borgne}, \& {Kneib}}]{r06}
{Richard}, J., {Pell{\'o}}, R., {Schaerer}, D., {Le Borgne}, J.-F., \& {Kneib},
  J.-P. 2006, \aap, 456, 861

\bibitem[{{Roche} {et~al.}(2003){Roche}, {Dunlop}, \& {Almaini}}]{roche03}
{Roche}, N.~D., {Dunlop}, J., \& {Almaini}, O. 2003, \mnras, 346, 803

\bibitem[{{Sawicki}(2002)}]{sawicki02}
{Sawicki}, M. 2002, \aj, 124, 3050

\bibitem[{{Sawicki} {et~al.}(2005){Sawicki}, {Stevenson}, {Barrientos},
  {Gladman}, {Mall{\'e}n-Ornelas}, \& {van den Bergh}}]{sawicki2005}
{Sawicki}, M., {Stevenson}, M., {Barrientos}, L.~F., {et~al.} 2005, \apj, 627,
  621

\bibitem[{{Schaerer}(2002)}]{schaerer2002}
{Schaerer}, D. 2002, \aap, 382, 28

\bibitem[{{Schaerer}(2003)}]{schaerer2003}
{Schaerer}, D. 2003, \aap, 397, 527

\bibitem[{{Schaerer} {et~al.}(2007){Schaerer}, {Hempel}, {Egami}, {Pell{\'o}},
  {Richard}, {Le Borgne}, {Kneib}, {Wise}, \& {Boone}}]{s07}
{Schaerer}, D., {Hempel}, A., {Egami}, E., {et~al.} 2007, \aap, 469, 47

\bibitem[{{Schaerer} \& {Pell{\'o}}(2005)}]{schaerer2005}
{Schaerer}, D. \& {Pell{\'o}}, R. 2005, \mnras, 362, 1054

\bibitem[{{Severgnini} {et~al.}(2006){Severgnini}, {Caccianiga}, {Braito},
  {Della Ceca}, {Maccacaro}, {Akiyama}, {Carrera}, {Ceballos}, {Page},
  {Saracco}, \& {Watson}}]{severgnini2006}
{Severgnini}, P., {Caccianiga}, A., {Braito}, V., {et~al.} 2006, \aap, 451, 859

\bibitem[{{Silva} {et~al.}(1998){Silva}, {Granato}, {Bressan}, \&
  {Danese}}]{silva1998}
{Silva}, L., {Granato}, G.~L., {Bressan}, A., \& {Danese}, L. 1998, \apj, 509,
  103

\bibitem[{{Simpson} {et~al.}(2006){Simpson}, {Almaini}, {Cirasuolo}, {Dunlop},
  {Foucaud}, {Hirst}, {Ivison}, {Page}, {Rawlings}, {Sekiguchi}, {Smail}, \&
  {Watson}}]{simpson2006}
{Simpson}, C., {Almaini}, O., {Cirasuolo}, M., {et~al.} 2006, \mnras, 373, L21

\bibitem[{{Smail} {et~al.}(1998){Smail}, {Ivison}, {Blain}, \&
  {Kneib}}]{smail1998}
{Smail}, I., {Ivison}, R.~J., {Blain}, A.~W., \& {Kneib}, J.-P. 1998, \apjl,
  507, L21

\bibitem[{{Smail} {et~al.}(2002){Smail}, {Owen}, {Morrison}, {Keel}, {Ivison},
  \& {Ledlow}}]{smail2002}
{Smail}, I., {Owen}, F.~N., {Morrison}, G.~E., {et~al.} 2002, \apj, 581, 844

\bibitem[{{Smith} {et~al.}(2002){Smith}, {Smail}, {Kneib}, {Czoske}, {Ebeling},
  {Edge}, {Pell{\'o}}, {Ivison}, {Packham}, \& {Le Borgne}}]{smith2002}
{Smith}, G.~P., {Smail}, I., {Kneib}, J.-P., {et~al.} 2002, \mnras, 330, 1

\bibitem[{{Somerville} {et~al.}(2004){Somerville}, {Moustakas}, {Mobasher},
  {Gardner}, {Cimatti}, {Conselice}, {Daddi}, {Dahlen}, {Dickinson},
  {Eisenhardt}, {Lotz}, {Papovich}, {Renzini}, \& {Stern}}]{somerville04}
{Somerville}, R.~S., {Moustakas}, L.~A., {Mobasher}, B., {et~al.} 2004, \apjl,
  600, L135

\bibitem[{{Somerville} {et~al.}(2001){Somerville}, {Primack}, \&
  {Faber}}]{somerville01}
{Somerville}, R.~S., {Primack}, J.~R., \& {Faber}, S.~M. 2001, \mnras, 320, 504

\bibitem[{{Stern} {et~al.}(2006){Stern}, {Chary}, {Eisenhardt}, \&
  {Moustakas}}]{stern2006}
{Stern}, D., {Chary}, R.-R., {Eisenhardt}, P.~R.~M., \& {Moustakas}, L.~A.
  2006, \aj, 132, 1405

\bibitem[{{Swinbank} {et~al.}(2007){Swinbank}, {Bower}, {Smith}, {Wilman},
  {Smail}, {Ellis}, {Morris}, \& {Kneib}}]{swinbank2007}
{Swinbank}, A.~M., {Bower}, R.~G., {Smith}, G.~P., {et~al.} 2007, \mnras, 72

\bibitem[{{Takata} {et~al.}(2003){Takata}, {Kashikawa}, {Nakanishi}, {Aoki},
  {Asai}, {Ebizuka}, {Inata}, {Iye}, {Kawabata}, {Kosugi}, {Ohyama}, {Okita},
  {Sasaki}, {Saito}, {Sekiguchi}, {Shimizu}, {Taguchi}, \&
  {Yoshida}}]{takata03}
{Takata}, T., {Kashikawa}, N., {Nakanishi}, K., {et~al.} 2003, \pasj, 55, 789

\bibitem[{{Tinsley} \& {Gunn}(1976)}]{tinsley76}
{Tinsley}, B.~M. \& {Gunn}, J.~E. 1976, \apj, 203, 52

\bibitem[{{Treu} {et~al.}(2005){Treu}, {Ellis}, {Liao}, {van Dokkum}, {Tozzi},
  {Coil}, {Newman}, {Cooper}, \& {Davis}}]{treu05}
{Treu}, T., {Ellis}, R.~S., {Liao}, T.~X., {et~al.} 2005, \apj, 633, 174

\bibitem[{{Werner} {et~al.}(2004){Werner}, {Roellig}, {Low}, {Rieke}, {Rieke},
  {Hoffmann}, {Young}, {Houck}, {Brandl}, {Fazio}, {Hora}, {Gehrz}, {Helou},
  {Soifer}, {Stauffer}, {Keene}, {Eisenhardt}, {Gallagher}, {Gautier}, {Irace},
  {Lawrence}, {Simmons}, {Van Cleve}, {Jura}, {Wright}, \&
  {Cruikshank}}]{Werner2004}
{Werner}, M.~W., {Roellig}, T.~L., {Low}, F.~J., {et~al.} 2004, \apjs, 154, 1

\bibitem[{{White} \& {Rees}(1978)}]{white78}
{White}, S.~D.~M. \& {Rees}, M.~J. 1978, \mnras, 183, 341

\bibitem[{{Wilson} {et~al.}(2004){Wilson}, {Huang}, {P{\'e}rez-Gonz{\'a}lez},
  {Egami}, {Ivison}, {Rigby}, {Alonso-Herrero}, {Barmby}, {Dole}, {Fazio}, {Le
  Floc'h}, {Papovich}, {Rigopoulou}, {Bai}, {Engelbracht}, {Frayer}, {Gordon},
  {Hines}, {Misselt}, {Miyazaki}, {Morrison}, {Rieke}, {Rieke}, \&
  {Surace}}]{wilson04}
{Wilson}, G., {Huang}, J.-S., {P{\'e}rez-Gonz{\'a}lez}, P.~G., {et~al.} 2004,
  \apjs, 154, 107

\bibitem[{{Yan} {et~al.}(2004){Yan}, {Dickinson}, {Eisenhardt}, {Ferguson},
  {Grogin}, {Paolillo}, {Chary}, {Casertano}, {Stern}, {Reach}, {Moustakas}, \&
  {Fall}}]{yan04}
{Yan}, H., {Dickinson}, M., {Eisenhardt}, P.~R.~M., {et~al.} 2004, \apj, 616,
  63

\bibitem[{{Yan} \& {Thompson}(2003)}]{yan03}
{Yan}, L. \& {Thompson}, D. 2003, \apj, 586, 765

\bibitem[{{Zheng} {et~al.}(1997){Zheng}, {Kriss}, {Telfer}, {Grimes}, \&
  {Davidsen}}]{zheng1997}
{Zheng}, W., {Kriss}, G.~A., {Telfer}, R.~C., {Grimes}, J.~P., \& {Davidsen},
  A.~F. 1997, \apj, 475, 469

\end{thebibliography}

\end{document}